\documentclass[twocolumn]{openjournal}

\usepackage[colorlinks=true, linkcolor=blue, citecolor=blue, urlcolor=blue]{hyperref}
\usepackage{amsmath}
\usepackage{natbib}
\usepackage{booktabs}
\usepackage{graphicx}
\usepackage{threeparttable}
\usepackage{orcidlink}

\newcommand{\temp}[2]{$T_{e,[\text{#1 #2}]}$}
\newcommand{\den}[2]{$n_{e,[\text{#1 #2}]}$}
\newcommand{\tempMD}{$T_{0}({\rm O}^{2+})$}
\newcommand{\te}{$T_{e}$ }
\newcommand{\dene}{$n_{e}$ }
\newcommand{\denoii}{$n_{e, \rm [O II]\lambda7325/\lambda3727}$}

\newcommand{\hii}{H\,\textsc{ii}}

\defcitealias{Delgado2023a}{MD23a}
\defcitealias{BPT1981}{BPT}
\defcitealias{Brazzini_2024}{B24}
\defcitealias{Delgado2023b}{MD23b}
\defcitealias{Vaught2023}{RV24}
\defcitealias{Nicholls2016}{N16}
\defcitealias{Morisset2016}{M16}

\newcommand{\revtwo}{}

\begin{document}

\title{Toward Unbiased Abundance Measurements in Inhomogeneous H \textsc{ii} Regions}
\shorttitle{Toward Unbiased Abundance Measurements in Inhomogeneous H \textsc{ii} Regions}
\author{Eric~Habjan$^{1,2}$\orcidlink{0009-0003-9547-0952}} 

\author{Christopher~Faesi$^{2}$\orcidlink{0000-0001-5310-467X}}

\author{Kathryn~Kreckel$^{3}$\orcidlink{0000-0001-6551-3091}}

\author{J.~Eduardo~M{\'e}ndez-Delgado$^{4}$\orcidlink{0000-0002-6972-6411}}

\author{Francesco~Belfiore$^{5,6}$\orcidlink{0000-0002-2545-5752}}

\author{Ryan~J.~Vaught$^{7}$\orcidlink{0000-0001-9719-4080}}

\author{Brent~Groves$^{8}$ \orcidlink{0000-0002-9768-0246}}

\author{Fabian~Scheuermann$^{3}$\orcidlink{0000-0003-2707-4678}}

\author{Thomas~G.~Williams$^{9}$\orcidlink{0000-0002-0012-2142}}

\author{Ralf~S.\ Klessen$^{10,11}$\orcidlink{0000-0002-0560-3172}}

\author{Amirnezam~Amiri$^{12,13}$\orcidlink{0000-0002-8553-1964}}

\author{Kathryn~Grasha$^{14}$\orcidlink{0000-0002-3247-5321}}

\author{Simon~Glover$^{10}$\orcidlink{0000-0001-6708-1317}}

\email{habjan.e@northeastern.edu}

\affiliation{$^{1}$Department of Physics, Northeastern University, 110 Forsyth St, Boston, MA 02115}
\affiliation{$^{2}$University of Connecticut, Department of Physics, 196A Auditorium Road, Unit 3046, Storrs, CT 06269, USA}
\affiliation{$^{3}$Astronomisches Rechen-Institut, Zentrum f\"{u}r Astronomie der Universit\"{a}t Heidelberg, M\"{o}nchhofstr. 12-14, D-69120 Heidelberg, Germany}
\affiliation{$^{4}$Universidad Nacional Aut\'onoma de M\'exico, Instituto de Astronom\'ia, AP 70-264, CDMX 04510, M\'exico}
\affiliation{$^{5}$European Southern Observatory, Karl-Schwarzschild Stra{\ss}e 2, D-85748 Garching bei M\"{u}nchen, Germany}
\affiliation{$^{6}$INAF -- Osservatorio Astrofisico di Arcetri, Largo E. Fermi 5, I-50157 Firenze, Italy}
\affiliation{$^{7}$Space Telescope Science Institute, 3700 San Martin Drive, Baltimore, MD 21218, USA}
\affiliation{$^{8}$International Centre for Radio Astronomy Research, University of Western Australia, 7 Fairway, Crawley, 6009 WA, Australia}
\affiliation{$^{9}$UK ALMA Regional Centre Node, Jodrell Bank Centre for Astrophysics, Department of Physics and Astronomy, The University of Manchester, Oxford Road, Manchester M13 9PL, UK}
\affiliation{$^{10}$Universit\"{a}t Heidelberg, Zentrum f\"{u}r Astronomie, Institut f\"{u}r Theoretische Astrophysik, Albert-Ueberle-Str.\ 2, 69120 Heidelberg, Germany}
\affiliation{$^{11}$Universit\"{a}t Heidelberg, Interdisziplin\"{a}res Zentrum f\"{u}r Wissenschaftliches Rechnen, Im Neuenheimer Feld 225, 69120 Heidelberg, Germany}
\affiliation{$^{12}$School of Astronomy, Institute for Research in Fundamental Sciences (IPM), Tehran, P.O. Box 19395-5531, Iran}
\affiliation{$^{13}$Department of Physics, University of Arkansas, 226 Physics Building, 825 West Dickson Street, Fayetteville, AR 72701, USA}
\affiliation{$^{14}$Research School of Astronomy and Astrophysics, Australian National University, Canberra, ACT 2611, Australia}
 
\begin{abstract}

Probing the chemical content of the interstellar medium (ISM) in nearby galaxies provides key insight into their chemical evolution and informs our interpretation of galaxies at higher redshift. However, nonlinear structure in the ISM, including density and temperature inhomogeneities, can bias chemical abundance measurements and systematically affect empirical calibrations derived from them. In this work, we investigate biases in $T_e$-derived oxygen abundance determinations and explore the physical properties that correlate with them. We combine $\mathrm{[O\,II]}\lambda\lambda3726, 3729$ measurements from SITELLE with a full suite of optical emission lines obtained with MUSE. From auroral emission lines ($\mathrm{[N\,II]}\lambda5755$, $\mathrm{[S\,III]}\lambda6312$, and $\mathrm{[O\,II]}\lambda\lambda7320, 7330$) and nebular emission lines (including $\mathrm{[N\,II]}\lambda6584$ and $\mathrm{[S\,III]}\lambda9069$), we derive electron densities, temperatures, and chemical abundances for a sample of H \textsc{ii} regions in five galaxies. We find that densities derived from the $\mathrm{[O\,II]}$ auroral-to-nebular ratio are $\sim10^3$ cm$^{-3}$, which is higher than the standard $\mathrm{[S\,II]}$ densities derived from nebular doublet ratios. We demonstrate that combining the $\mathrm{[N,II]}$ electron temperature with the density inferred from the $\mathrm{[O\,II]}$ auroral-to-nebular line ratio yields singly ionized oxygen abundances consistent with literature expectations for a prescription insensitive to density inhomogeneities. We also find that the $\mathrm{[S\,III]}$ temperature provides a reliable estimate of $T_{e,\mathrm{[O\,III]}}$, enabling robust measurements of doubly ionized oxygen abundances. Overall, these results indicate that the abundance discrepancy factor could be higher in more chemically evolved H \textsc{ii} regions.

\end{abstract}


\section{Introduction}
\label{sec:intro}

The gas-phase chemical abundance, metallicity, of the interstellar medium (ISM) is an important tracer of galaxy and the chemical evolution. Metallicity is estimated by taking the ratio of emission lines. \hii\ regions often have bright emission lines due to the O and B type stars \citep{Osterbrock} that ionize the surrounding gas. Given the short lifetime of these stars, the metallicity of an \hii\ region acts as a tracer for the present-day chemical \revtwo{composition} of the \revtwo{ISM}. Metallicity is a crucial parameter in stellar, galactic and cosmic chemical evolution models \citep[e.g.,][]{Pettini2004, Kewley2002, Kobulunicky2004}, hence accurate measurements of metallicity are pivotal for theoretical modeling in astrophysics. 

\revtwo{Elemental abundances in ionized nebulae are commonly derived using either recombination lines (RLs) or collisionally excited lines (CELs). RLs are produced when ions recombine with free electrons, while CELs arise from collisional excitation followed by radiative de-excitation. However, abundances derived from RLs and CELs have long been known to disagree, with RLs typically yielding higher metallicities than CELs} \citep{Wyse_1942, Peimbert1967, Rubin_1986, Rodriguez_2010, Nicholls_2012, Peimbert_2017, Garcia-Rojas_2020, Delgado2023a}. The discrepancy between the abundances derived from CELs and RLs is quantified by the abundance discrepancy factor (ADF; see, e.g., \cite{Garcia-Rojas_2007}). Studies \citep{Peimbert1967, Peimbert_1969, Binette2012, Peimbert_2013, Berg_2015, Nicholls_2020, Berg_2020, Delgado2023b} have shown that density and temperature inhomogeneities may be present within the different ionization zones of \hii\ regions, and are the most likely cause of the ADF. In particular, CELs are exponentially dependent on temperature while RLs are roughly proportional to electron density and inverse electron temperature \citep{Osterbrock}. Thus, when there are clumps of gas within an \hii\ region with different temperatures, the ratio of emission lines tend to be biased to clumps with higher temperatures \citep{Delgado2023a}. These systematically larger electron temperatures from CELs subsequently cause underestimated metallicities in comparison to RL-derived metallicities (i.e., the ADF). 

The ADF has been a long-standing problem in metallicity derivations \citep{Peimbert1967, Peimbert1971, Jamet_2005, Delgado2023a, Delgado2023b, Vaught2023}. To address this problem, \citet[hereafter MD23a]{Delgado2023a} presented an empirical method that relates the $\rm [N\, II]$ electron temperature to a RL-derived metallicity (see Section \ref{subsec:empirical_temp_methods}). In this work, we aim to study the density and temperature prescriptions that yield a metallicity that best agrees with this RL-derived empirical method. This will inform what physical quantities are most affected by inhomogeneities in the gas, thus minimizing the ADF. 

An additional difficulty in metallicity measurement with either CELs or RLs is that both methods require very faint emission lines \revtwo{(e.g., $\rm [N\, II]\lambda$~5755, $\mathrm{O\, II}$ V1 RLs)}. This has lead to the development of `strong line' calibrations being employed to estimate metallicity \revtwo{\citep{Kewley2002, Kobulunicky2004, P05, Dopita2016, Pilyugin2016, RosalesOrtega2026}}. These strong line methods are calibrated using metallicities obtained through the `direct $T_{e}$' method \citep{PP04, Marino2013, Pilyugin2016, P05, Delgado2023a, RosalesOrtega2026}, theoretical models \citep{M91, Kewley2002, Kobulunicky2004, T04} or a combination of both \citep{Z94, Dopita2016, D02}. Different strong-line metallicity calibrations can show variations of $\sim$1 dex in metallicities \citep[e.g., ][]{Kewley2008, Teimoorinia2021} derived for the same objects (e.g., \hii\ regions, emission-line galaxies). 

With an increasing number of spectroscopic studies being done in both nearby and in high-redshift galaxies, many different direct and strong line methods are being employed to estimate metallicity \revtwo{\citep{Maiolino2019,Kewley2019}}. \revtwo{Understanding the ADF is a necessary first step toward accurately tracing metallicity evolution across galactic and cosmological scales.} In this work, we aim to prescribe a direct method as well as a strong line method that produces the most accurate estimation of metallicity. Specifically, this is done by comparing CEL derived abundances with the RL derived relation from \citetalias{Delgado2023a}. Such a comparison will demonstrate which electron density and temperature diagnostics are most affected by the density/temperature inhomogeneities that give rise to the ADF. 

In \cite{Habjan2026a}, observations of five nearby galaxies \revtwo{(NGC~628, NGC~2835, NGC~3351, NGC~3627, and NGC~4535)} from Spectromètre Imageur à Transformée de Fourier pour l'Etude en Long et en Large de raies d'Emission \revtwo{\citep[SITELLE; see, e.g.,][]{Martin2016, Rousseau-Nepton2018}} are analyzed to obtain blended $\rm[O\, II]\lambda\lambda$3726,3729 (hereafter $\rm[O\, II]\lambda$3727) nebular line fluxes. We pair these $\rm[O\, II]\lambda$3727 fluxes with Multi-Unit Spectroscopic Explorer (MUSE) data from the Physics at High Angular Resolution in Nearby GalaxieS (PHANGS) survey \citep{Emsellem2022}. The suite of emission lines from MUSE and SITELLE allow for direct \te metallicity calculations in \hii\ regions that have spatial resolution of $\sim$100~pc. We compare our derived metallicities to those calculated using the RL-derived calibration derived by \citetalias{Delgado2023a}. 

The paper is organized as follows: the observations and auroral line fitting are described in Section \ref{sec:data}. Electron temperatures and densities are derived and presented in Section \ref{sec:Electron Densities and Temperatures}. Oxygen and nitrogen abundances are calculated and studied in Section \ref{sec:abundances}. The implications of these results are outlined and explored in Section \ref{sec:discussion} and we conclude in Section \ref{sec:conclusion}. 

\section{Data}
\label{sec:data}

Our analysis focuses on a sample of five galaxies (Table \ref{tbl:sample_observations}), observed by SITELLE \citep{Grandmont2012SPIE.8446E..0UG} at the Canada-France-Hawaii Telescope (CFHT); two of these galaxies (NGC~2835, NGC~4535) were observed by proposal ID 20AF06 (PI: Hughes) and three (NGC~628, NGC~3351, NGC~3627) by the Star formation, Ionized Gas, and Nebular Abundances Legacy Survey \citep[SIGNALS; ][proposal ID 20BP41; PI:  Rousseau-Nepton]{Rousseau-Nepton2018}. In this section, we summarize the data and catalogs used to identify \hii\ regions in these galaxies, drawing from both the MUSE and SITELLE instruments. 

\begin{table*}
\caption{General properties for galaxies in our sample.}
\footnotesize
\centering
%\hspace*{-0.125\textwidth}
\begin{tabular}{@{}cccccccc@{}}
\toprule
Galaxy & RA$^{1}$ & Dec$^{1}$ & Distance$^{2}$ & $v_{\rm sys}$$^{3}$ & log$_{10} M_{\star}$$^{4}$ & Log(SFR)$^{4}$ \\
Name & [deg] & [deg] & [Mpc] & [km s$^{-1}$] & [M$_{\odot}$] & [M$_{\odot} {\rm yr}^{-1}$] \\
\midrule
NGC~628 & 24.173855 & 15.783643 & 9.84 & 651 & 10.34 & 0.24 \\
NGC~2835 & 139.47044 & -22.35468 & 12.2 & 867 & 10.00 & 0.09 \\
NGC~3351 & 160.99065 & 11.70367 & 10.0 & 775 & 10.36 & 0.12 \\
NGC~3627 & 170.06252 & 12.9915 & 11.3 & 715 & 10.83 & 0.58 \\
NGC~4535 & 188.5846 & 8.197973 & 15.8 & 1954 & 10.53 & 0.33 \\
\bottomrule
\end{tabular}
\begin{tablenotes}
\centering
\small
\item \textbf{Notes.} $^{1}$From \cite{Santoro2022}. $^{2}$From \cite{Anand2021}.
 
\item $^{3}$From \cite{Makarov2014}. $^{4}$From \cite{Leroy2021}.
\end{tablenotes}
\label{tbl:sample_observations}
\end{table*}

\subsection{PHANGS-MUSE Data}
\label{subsec:nebcatalog}

The Multi Unit Spectroscopic Explorer (MUSE) is a panoramic integral-field spectrograph at the Very Large Telescope. MUSE is able to achieve a resolution of $<100$ pc for nearby ($<$20 Mpc) galaxies, enabling \hii\ regions to be resolved on the inter-cloud scale. The PHANGS-MUSE Nebular Catalog (hereafter Nebular Catalog) constructed by \cite{Groves2023} uses H$\alpha$ morphology from the PHANGS-MUSE sample as well as standard \revtwo{Baldwin-Phillips-Terlevich \citep[BPT;][]{BPT1981} diagnostics to identify \hii\ regions. In particular, regions that fall below the \cite{Kauffmann2003} diagnostic curve in the $\rm [O\, III]\, /\, H\beta$ versus $\rm [N\, II]\, /\, H\alpha$  diagram and below the \cite{Kewley2006} diagnostic curve in the $\rm [O\, III]\, /\, H\beta$ versus $\rm [S\, II]\, /\, H\alpha$  are flagged as \hii\ regions, resulting in 23,244 \hii\ regions across 19 galaxies.} Each of the five galaxies observed with SITELLE were observed as part of the PHANGS-MUSE survey and have \hii\ regions tabulated in the Nebular Catalog. The general properties of our sample are shown in Table \ref{tbl:sample_observations}. We use the Baldwin-Phillips-Terlevich \citep{BPT1981} flags from the Nebular Catalog to identify 5,658 \hii\ regions in our sample. 

\subsection{SITELLE Data}
\label{subsec:sitelle_data}

SITELLE is an optical Fourier Transform spectrograph with a field of view of 11 arcmin x 11 arcmin, large enough to contain the full star-forming disk of each galaxy. This work makes use of data obtained using SITELLE's SN1 filter that covers 3650-3850~{\AA} and includes the $\rm[O\,II]\lambda$3727 emission line. \cite{Habjan2026a} describes the data reduction and emission line flux measurements corresponding to the Nebular Catalog footprints. Of the full set of nebulae covered by the SITELLE data, 743 (8.4\%) are detected in $\rm[O\,II]$ with S/N$>$3. Based on the \citetalias{BPT1981} cuts described above, 680 of these are classified as \hii\ regions. Each of these line fluxes is corrected for both foreground and internal extinction, applying the same E(B-V) reddening and \cite{ODonnnell1994} extinction law as was obtained from the Balmer decrement in the MUSE data.

\subsection{Auroral Line Fitting}
\label{subsec:auroral}

The MUSE Data Analysis Pipeline (DAP) described in \cite{Emsellem2022} provides a method based on the gist code \citep{Bittner2019A&A...628A.117B}, to use penalised PiXel-Fitting (pPXF) \citep{Cappellari2004PASP..116..138C} to fit stellar and emission line properties of the spectra from the MUSE observations. However, after stellar background continuum subtraction, a residual continuum may be present at the location of faint lines, which if ignored, may bias their measured intensity given their intrinsic faint flux. In order to more accurately measure auroral line fluxes in our \hii\ region sample, we refit the $\rm[N\,II]\lambda$5755, $\rm[S\,III]\lambda$6312, and $\rm[O\,II]\lambda\lambda$7320,7330 (hereafter $\rm[O\,II]\lambda$7325) auroral lines. This allows for line measurement uncertainties and quality cuts to be made consistently with the $\rm[O\,II]$ SITELLE fluxes while also providing a direct comparison to the pPXF procedure used in \cite[][hereafter B24]{Brazzini_2024}.

\begin{figure*}
\centering
  \includegraphics[width=\textwidth]{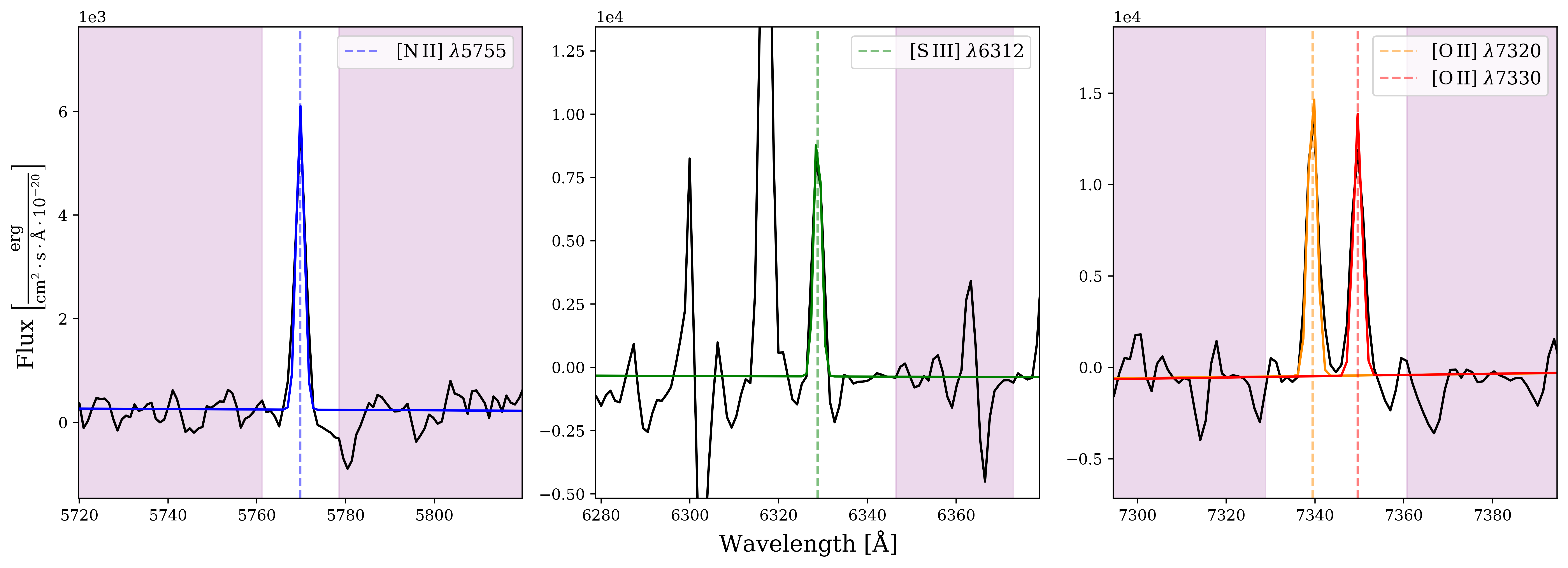}
  \caption{Example fits of each auroral line from Region 144 in NGC~2835; $\rm [N\, II]\lambda$5755 in blue (S/N $\sim$ 18), $\rm [S\, III]\lambda$6312 in green (S/N $\sim$ 12), $\rm [O\, II]\lambda$7320 in orange (S/N $\sim$ 14), and $\rm [O\, II]\lambda$7330 in red (S/N $\sim$ 13). The MUSE DAP nebular spectrum is shown as the black line and each panel shows the fits for each of the four auroral lines in our spectral range. Each of these fits were done using a Gaussian with an added linear polynomial to account for residual background continuum. The noise for each fit is calculated using the spectral background regions discussed in Appendix \ref{appendix:background}; the overlapping spectral regions from which noise is extracted are shaded in purple.}
  \label{fig:auroral_lines}
\end{figure*}

This is done by fitting each auroral line with a Gaussian and modeling the surrounding continuum (see Appendix \ref{appendix:background} for exact wavelength ranges) with a straight line. \revtwo{To model the continuum, which is approximately flat immediately surrounding each auroral line, we first fit a first-order polynomial to continuum regions on either side of the line. The resulting continuum model is then subtracted from the spectrum, and the auroral line is modeled with a Gaussian.} Each of the four auroral lines are fitted using the \texttt{curve$\_$fit} method from \texttt{SciPy} \citep{scipy}. The amplitude, central redshifted wavelength and velocity dispersion are taken as free parameters during the fitting process. An initial guess for the central redshifted wavelength for singly ionized auroral lines (i.e. N$^{+}$ and O$^{+}$) is made using the $\rm[N\,II]\lambda$6584 velocity from the Nebular Catalog. For the doubly ionized auroral lines (i.e. S$^{2+}$) we use the $\rm[O\,III]\lambda$5007 velocity to estimate the central redshifted wavelength. Similarly, we use the $\rm[N\,II]\lambda$6584 velocity dispersion from the Nebular Catalog to estimate the $\rm[N\,II]\lambda$5755 and $\rm[O\,II]\lambda$7325 Gaussian widths while we use the $\rm[O\,III]\lambda$5007 velocity dispersion to estimate the width of $\rm[S\,III]\lambda$6312. The estimated central redshifted wavelength was given $\pm$0.5 \AA\ of freedom from the initial estimate as the velocities from the Nebular catalog are very accurate. \revtwo{The standard deviation of the Gaussian $\sigma_\lambda$ could vary in the range $\sigma_{\rm \lambda, SL}$~$<$ $\sigma_\lambda$~$<$~$\sigma_{\rm \lambda, SL} + 0.5$ [\AA] where $\sigma_{\rm \lambda, SL} = \lambda_{\rm rest} \times (\sigma_{v, \rm SL} / c)$, $\lambda_{\rm rest}$ is the auroral line rest wavelength, and $\sigma_{v, \rm SL}$ is the strong line velocity dispersion from the Nebular Catalog.} This restriction was implemented for $\sigma$ as the auroral line velocity dispersions have a tendency to be larger than the strong line velocity dispersions, but we also did not want to include any emission features that are unphysically broad. Examples of each auroral line fit are shown in Figure \ref{fig:auroral_lines}. 

We determine uncertainties using a \revtwo{Monte Carlo (MC) method described below}. A Gaussian noise distribution is created with a mean of zero and a standard deviation equal to that of the surrounding continuum (see Appendix \ref{appendix:background}). Random draws from this distribution are made and added to the spectrum. Our fitting process is carried out 1,000 times with added Gaussian noise, and the standard deviation in the obtained auroral line fluxes is taken to be the 1 $\sigma$ uncertainty. We correct for extinction using the same process described in Section \ref{subsec:sitelle_data}. 

In order to minimize the effects of sky contamination in our auroral line fluxes, we only consider auroral line fits that have signal-to-noise (S/N) $>$ 5. The effect of sky contamination for the galaxies in our sample affects $\rm [N\,II]\lambda$5755 to a smaller extent, however the spectral ranges near the redshifted features of $\rm [S\,III]\lambda$6312 and $\rm [O\,II]\lambda$7325 show signs of contamination. \revtwo{The $\rm [O\,II]\lambda7320 / 
\lambda7330$ ratio is not strictly independent of physical conditions, since the transitions originate from levels with different collisional de-excitation rates; however, the extremely high critical densities involved ensure that it remains largely insensitive to density variations up to $n_e \approx 10^{5.5}\, \rm cm^{-3}$.} As a quality check, we find that the $\rm [O\,II]\lambda$7320/$\lambda$7330 ratio measured for our sample is 1.27 $\pm$ 0.03. This is in good agreement with the $\rm [O\,II]\lambda$7320/$\lambda$7330 ratio found \revtwo{by} Section 7.1.1 of \citet[][hereafter RV24]{Vaught2023} and the theoretical predictions from \cite{Kisielius2009} of 1.24 at densities \revtwo{lower than $n_e\approx 10^{5.5}\, \rm cm^{-3}$}. To check the quality of our $\rm [S\,III]\lambda$6312 fluxes, we compare our \te derived from $\rm [S\,III]\lambda$6312 to calibrations from the literature in Section \ref{subsec:siii_te_quality_check}. As a last step, we visually inspected each auroral line fit in our sample and found unexpectedly broad auroral lines in some regions. Since we are using the velocity dispersion of CELs to estimate the width of each Gaussian fit, any broadening due to movement of the gas should already be accounted for. Thus, auroral line fluxes in which the observed broadening far exceeds the upper bound of our allowed Gaussian widths are considered to be contaminated by sky line emission and removed from our catalog. 

Lastly, we find strong agreement with \citetalias{Brazzini_2024} for nearly all auroral line measurements, with differences typically below 10\%. The main exceptions are a small number of $\rm [N\, II]\lambda$5755 measurements, all of which occur in regions that only marginally satisfy our S/N $> 5$ criterion and are located in NGC~3627. These spectra often exhibit an absorption feature redward of the galaxy-frame $\rm [N\, II]\lambda$5755 line, which may complicate the stellar continuum subtraction and bias the measured auroral flux. We therefore exclude these regions from our analysis because their measured fluxes differ by $\sim$40\% from those obtained using the fitting procedure of \citetalias{Brazzini_2024}. The origin of this discrepancy remains unclear \revtwo{and will be subject to future analysis}.

\section{Electron Densities and Temperatures}
\label{sec:Electron Densities and Temperatures}

In this section, we describe the derivation of electron densities \dene and temperatures $T_{e}$ using the extinction corrected emission line fluxes discussed in Section \ref{sec:data}. The methods for each quantity are described first and then followed by the results.

\subsection{Electron Density and Temperature}
\label{subsec:temp_den}

\subsubsection{\texorpdfstring{\te and \dene}{Te and ne}Methodology}
\label{subsec:temp_den_methods}

We first derive the $\rm [S\, II]$ density \den{S}{II} using the $\rm [S\, II]\lambda$6731/$\lambda$6716 diagnostic. This density is then combined with the $\rm [O\, II]\lambda$7325/$\lambda$3727, $\rm [N\, II] \lambda5755 / \lambda6584$, and $\rm [S\, III] \lambda6312 / \lambda9069$ temperature sensitive line ratios to derive the average electron temperature of each ion using the \texttt{getCrossTemDen} method from PyNeb \citep{pyneb}. MC uncertainties are calculated on each \te and \den{S}{II} using $10^{3}$ Gaussian iterations by propagating flux uncertainties. Each temperature derivation yields a unique \den{S}{II}, thus we take \den{S}{II} to be a weighted average; the weight is equal to the inverse square of the 1 $\sigma$ uncertainty of each independently derived \den{S}{II}. 

\begin{figure*}
\centering
  \includegraphics[width=0.7\textwidth]{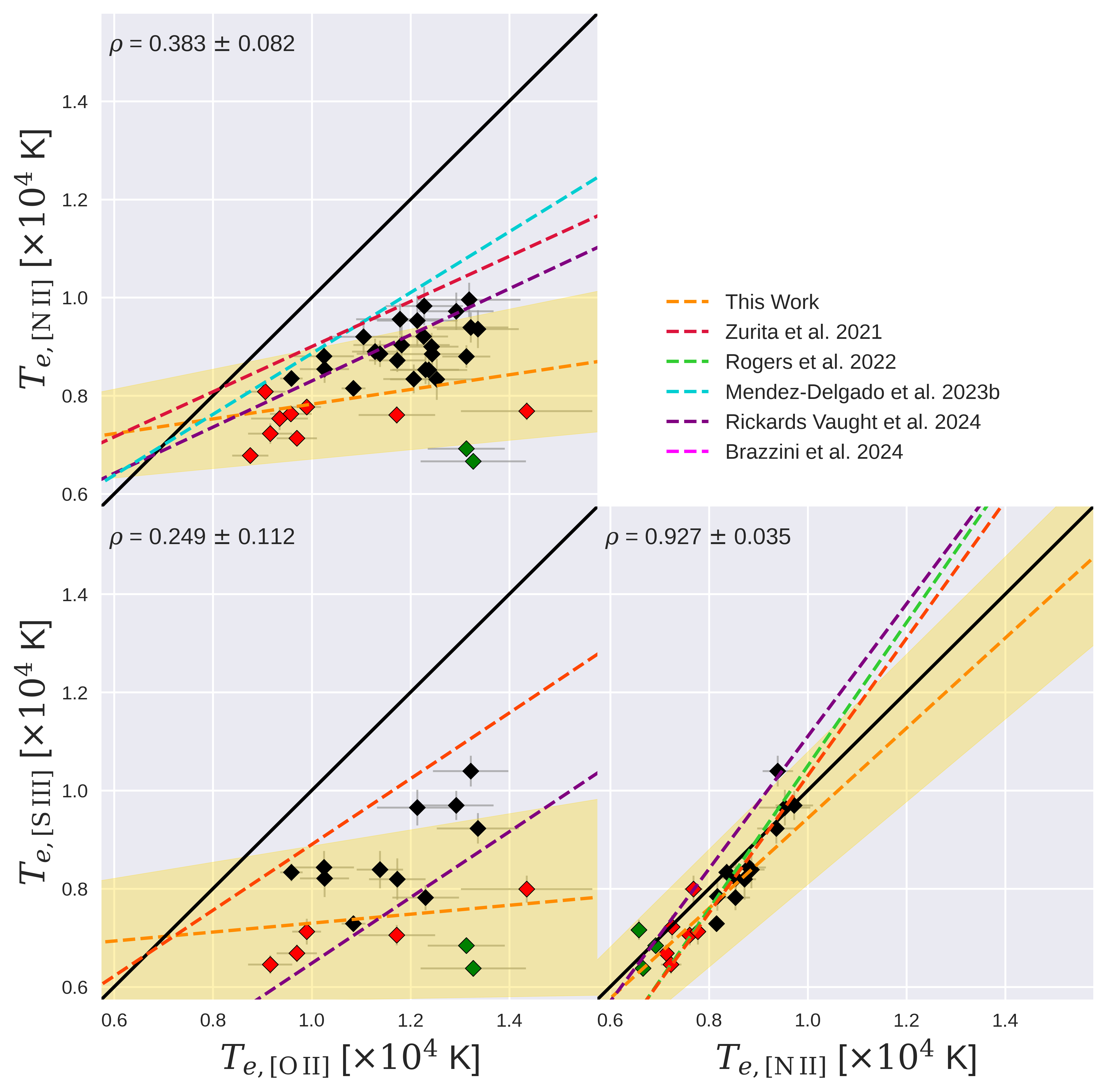}
  \caption{Each of our directly measured electron temperatures plotted against each other. NGC~628 temperatures are shown in red, NGC~2835 in black and NGC4535 in green. There are no detections for NGC~3351 and NGC~3627. The dashed gold lines are the WLS fits to the \te -- \te relationships, which are shown in Table \ref{tbl:temptemp}. The 1 $\sigma$ uncertainty is shaded in gold around each WLS fit. The 1-1 line is plotted in black to aid in visual comparison, and the Spearman rank correlation coefficient for each \te -- \te relationship is shown in upper left corner of each plot. The respective trend lines from different studies are plotted in each \te -- \te relation with \cite{Zurita2021} in dark red, \cite{Rogers2022} in lime green, \citetalias{Delgado2023b} in turquoise, \citetalias{Vaught2023} in purple, and \citetalias{Brazzini_2024} in fuchsia.}
  \label{fig:temp_corner}
\end{figure*}

We present our \temp{N}{II}, \temp{O}{II} and \temp{S}{III} in Figure \ref{fig:temp_corner}. We fit a weighted least-squares (WLS) line to each \te--\te\ relationship
\revtwo{using inverse-variance weights derived from the MC uncertainties.
We also calculate the Spearman rank correlation coefficient $\rho$.} We propagate uncertainties from each \te to calculate the uncertainties in the WLS fit and the Spearman rank correlation coefficient. The fitted \te -- \te relationships are presented in Table \ref{tbl:temptemp}. 

\begin{table*}
\centering
\caption{\te -- \te  Weighted least squares relationships, \\ \revtwo{with all temperatures expressed in units of $10^4$~K}.}
\begin{tabular}{cc}
\toprule
WLS relationship & Number of regions \\
\midrule

\temp{N}{II} = 0.15($\pm$ 0.05) $\times$ \temp{O}{II} + 0.63($\pm$ 0.15) & 35 \\

\temp{S}{III} = 0.09($\pm$ 0.08) $\times$ \temp{O}{II} + 0.67($\pm$ 0.07) & 18\\

\temp{S}{III} = 0.92($\pm$ 0.07) $\times$ \temp{N}{II} + 0.03($\pm$ 0.06) & 21\\

\bottomrule
\end{tabular}
\label{tbl:temptemp}
\end{table*}

\begin{figure}
\centering
  \includegraphics[width=0.4\textwidth]{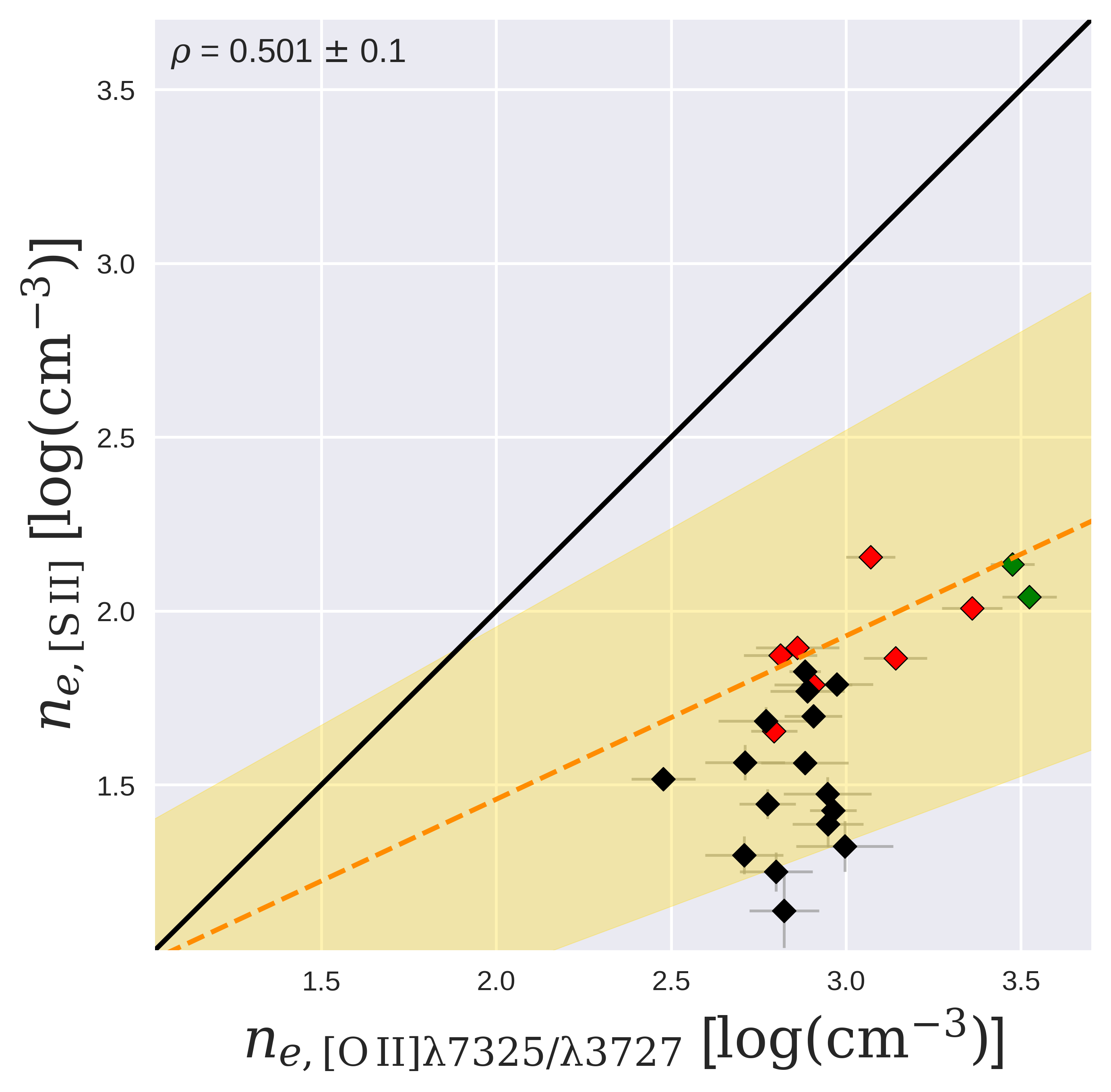}
  \caption{\den{S}{II} plotted against \denoii. NGC~628 is shown in red, NGC~2835 in black, and NGC~4535 in green. There are no detections for NGC~3351 and NGC~3627. The WLS fit is in gold and the shaded region represents the 1 $\sigma$ uncertainty. The Spearman rank correlation coefficient $\rho$ is shown in the top left corner and a 1-1 line is plotted in black.}
  \label{fig:den-den}
\end{figure}

\citet[][hereafter MD23b]{Delgado2023b} found that density inhomogeneities may be the cause of an underestimation of \dene when the $\rm [S\, II]\lambda$6731/$\lambda$6716 diagnostic is used. Thus, a \dene is calculated using the $\rm [O\, II]\lambda$7325/$\lambda$3727 diagnostic to derive an $\rm [O\, II]$ density \denoii. The diagnostic $\rm [O\, II]\lambda$7235/$\lambda$3727 is commonly used to derive an \temp{O}{II}, but it also has a strong density sensitivity in the range $10^{2}$ cm$^{-3}$ $<$ \dene $<$ $10^{6}$ cm$^{-3}$ \citepalias{Delgado2023b}. In this density range, the diagnostic $[\text{NII}]\,\lambda5755\slash\lambda6584$ is approximately constant. We derive \denoii\ using the $\rm [O\, II]$ and $\rm [N\, II]$ auroral-to-nebular line ratios with the implicit assumption that \temp{N}{II} = \temp{O}{II}. This assumption is a typical prediction of simple photoionization models of typical star-forming regions \citep{Morisset_2015, ValeAsari_2016, Ferland_2017, Delgado2023b, Vaught2023, Cataldi:25}. We propagate 1 $\sigma$ uncertainties to calculate MC uncertainties for \denoii. We make a direct comparison of \den{S}{II} and \denoii\ in Figure \ref{fig:den-den}. A WLS line is fitted and the Spearman rank correlation coefficient $\rho$ is calculated for this relationship.

Both the $\rm [S\, II]\lambda$6731/$\lambda$6716 and $\rm [O\, II]\lambda$7325/$\lambda$3727 diagnostics are insensitive to density when \dene $<$ $10^{2}$ cm$^{-3}$. For this reason, when \dene is in the low density limit, we set \dene equal to 100 cm$^{-3}$ $\pm$ 100 cm$^{-3}$ when deriving ionic abundances in Sections \ref{subsec:OII_abundances}, \ref{subsec:metal} and \ref{subsec:nitrogen}. 

\subsubsection{\texorpdfstring{\dene}{ne}Results}
\label{subsec:den_results}

The analysis described in Section \ref{subsec:temp_den_methods} yielded a total of 176 \den{S}{II} values with S/N $>$ 3 across the sample of 5,658 \hii\ regions in five galaxies. Of these 176 \den{S}{II} detections, only 15 \hii\ regions had \den{S}{II} $>$ 100 cm$^{-3}$ (i.e., not in the low density limit), thus we set the majority of \den{S}{II} detections equal to 100 $\pm$ 100 cm$^{-3}$ for ionic abundance determinations. Conversely, there are 27 \denoii\ detections across the entire sample, and none of these detections are in the low density limit; the \denoii\ detections have an average value of 1060 $\pm$ 46 cm$^{-3}$.

The \hii\ regions that have a detection for both \den{S}{II} and \denoii\ are shown in Figure \ref{fig:den-den}. There is a $\sim\!10^{3}$ cm$^{-3}$ difference in the measured density from $\rm [O\, II]\lambda$7325/$\lambda$3727 and $\rm [S\, II]\lambda$6731/$\lambda$6716. In \citetalias{Delgado2023b}, it is found that the average density difference between these diagnostics is $\sim\!300$ cm$^{-3}$. Furthermore, \citetalias{Delgado2023b} finds that when \den{S}{II} $>$ $10^{3}$ cm$^{-3}$, then nebular-to-nebular density diagnostics are in good agreement with densities derived using auroral-to-nebular diagnostics. The majority of \den{S}{II} detections by \citetalias{Delgado2023b} are not in the low density limit (i.e., \den{S}{II} $>$ $10^{2}$ cm$^{-3}$), which may explain why the ratio of \den{S}{II} and \denoii\ is a factor of $\sim$3 larger than the density offset presented in \citetalias{Delgado2023b}. 

Another potential explanation for this factor of $\sim$3 difference could be diffuse ionized gas (DIG) contribution.
\revtwo{In general, DIG is spatially extended beyond the typically defined \hii\ region boundary and can bias low-ionization emission lines, particularly in low-surface-brightness \hii\ regions. For example, \cite{Zhang2017} show that DIG contamination can affect lines such as $\rm [N\,II]\lambda6548$ and $\rm [O\,II]\lambda3727$. In \citetalias{Delgado2023b}, the spectra used are slit observations focused on bright \hii\ regions with large contrast relative to the DIG. Since the abundance derivations in this work require auroral line detections and are bright in $\rm H\alpha$ emission, the DIG most likely does not significantly affect our measurements. In a future work, a DIG subtraction should be carried out to quantify the bias induced from the DIG and from the presence of density inhomogeneities.}

\subsubsection{\texorpdfstring{\te}{} \hspace{-3mm} Results}
\label{subsec:temp_results}

The methods described in Section \ref{subsec:temp_den_methods} produced 53 \temp{O}{II}, 202 \temp{N}{II} and 27 \temp{S}{III} detections. We see in Figure \ref{fig:temp_corner} that \temp{N}{II} and \temp{S}{III} follow a nearly linear relationship, where \temp{S}{III} $\sim$ 0.95 $\times$ \temp{N}{II} with a Spearman rank correlation coefficient $\rho$ of 0.93 $\pm$ 0.04. The global trends found in \cite{Rogers2022}, \cite{Zurita2021} and \citetalias{Vaught2023} show a steeper slope and superlinear relationship for \temp{N}{II} -- \temp{S}{III}. Each of the trend lines from other studies in Figure \ref{fig:temp_corner} agree to within 1-$\sigma$ between the range of $0.6 \times 10^4$ K and $0.9 \times 10^4$ K, indicating that this \te -- \te relationship agrees well with other studies. In particular, the \temp{N}{II} and \temp{S}{III} relationship from Figure 9 of \citetalias{Vaught2023} shows that across their entire sample that \temp{S}{III} is often larger than \temp{N}{II}. However, in NGC~2835 and NGC~628, the majority of \hii\ regions from \citetalias{Vaught2023} show that \temp{N}{II} $\sim$ \temp{S}{III}, which can be seen in Figure \ref{fig:temp_corner}. It should be noted that although \citetalias{Vaught2023} makes use of the same MUSE observations, the \hii\ region boundaries are defined differently. Regardless, these findings indicate that the obtained \temp{N}{II} -- \temp{S}{III} relationship is strongly dependent on the galaxies in the sample. 

We find that \temp{O}{II} is only weakly correlated with \temp{N}{II} and \temp{S}{III}, with $\rho$ of $0.38 \pm 0.09$ and $0.25 \pm 0.11$, respectively. We find an average \temp{O}{II} of $1.25 \times 10^{4}$\,K with a standard deviation of $0.20 \times 10^{4}$\,K. The large values of \temp{O}{II} and the lack of correlation between \temp{O}{II} with other \te in Figure \ref{fig:temp_corner} can be explained by either the density dependence of the $\rm [O\, II]\lambda$7325/$\lambda$3727 temperature diagnostic or contribution from the DIG (see Section \ref{subsec:den_results}). \revtwo{The weak correlation between \temp{O}{II} and the other \te shown in Figure \ref{fig:temp_corner} is contrary to what is found in \citetalias{Delgado2023b}, which shows that \temp{O}{II}, despite being overestimated due to density inhomogeneities, is still correlated with \temp{N}{II}.} \revtwo{The lack of correlation between \temp{O}{II} and other \te in this work} may be evidence that the DIG \revtwo{or another mechanism} is biasing \temp{O}{II} measurements. \revtwo{Furthermore, the relations between \temp{O}{II} and other \te\ in this work are found to be shallower than the relations found by \citetalias{Vaught2023}, \citetalias{Delgado2023b}, \cite{Rogers2022}, and \cite{Zurita2021}. Some of this variation may reflect differences in the galaxy samples used to derive each relation, as sample-dependent differences in metallicity, ionization conditions, and star formation environments can affect the observed \te--\te\ trends. However, as shown by \citetalias{Vaught2023}, for example, both the \temp{N}{II}--\temp{O}{II} and the \temp{S}{III}--\temp{O}{II} relations exhibit substantial scatter, suggesting an underlying mechanism that drives departures from a simple linear relationship.}

\subsection{Empirical Electron Temperatures}\label{subsec:empirical_temperature}

\subsubsection{Empirical \texorpdfstring{\te}{} \hspace{-3mm} Methodology}
\label{subsec:empirical_temp_methods}

Neither the MUSE nor SITELLE spectral range covers the $\rm [O\, III]\lambda$4363 auroral line. However, because \temp{O}{III} is widely used in the literature we estimate it using an empirical calibration\revtwo{. In particular, we use the relation from \citetalias{Brazzini_2024}:}

\begin{gather}\begin{aligned}\label{eq:oiii_Te_B24}
\text{\temp{O}{III}} = (0.80 \pm 0.02) \times \text{\temp{S}{III}} + (0.20 \pm 0.02),
\end{aligned}\raisetag{0\baselineskip}
\end{gather}  

\noindent \revtwo{in units of $10^4$~K.} \citetalias{Brazzini_2024} calibrates this relation using \hii\ region spectra obtained from \cite{Guseva_2011} and CHAOS  \citep{Berg_2020, Rogers2022}.

\revtwo{An additional empirical relation is used to find the temperature at which RLs and CELs yield the same electron temperature. This temperature was first formalized in \cite{Peimbert1967} and was then measured in \citetalias{Delgado2023a}. To calibrate this relation}, \citetalias{Delgado2023a} used spatially resolved \hii\ regions with deep spectra to measure the variations in gas temperature for the O$^{2+}$ ion using both RLs and CELs. \revtwo{This temperature, \tempMD\, was found by comparing \temp{O}{III} and $T_{e, \text{O II}}$ at different ionization volumes throughout various \hii\ regions using Equation~1 in \citetalias{Delgado2023a}. The calibration between \tempMD\ and \temp{N}{II} is:}

\begin{gather}\begin{aligned}\label{eq:oiii_Te_MD23}
\text{\tempMD} = 1.17 \times \text{\temp{N}{II}} - 3340,
\end{aligned}\raisetag{0\baselineskip}
\end{gather}

\noindent \revtwo{in units of $10^4$~K and where \temp{N}{II} was derived using auroral-to-nebular line ratios in \citetalias{Delgado2023a}.} Uncertainties on \temp{O}{III} and \tempMD\ are calculated by employing the MC method described in Section \ref{subsec:temp_den_methods}. 

\subsubsection{Empirical \texorpdfstring{\te}{} \hspace{-3mm} Results}
\label{subsec:empirical_temp_results}

We calculate \temp{O}{III} for all 27 \temp{S}{III} detections in our sample using Equation \ref{eq:oiii_Te_B24}. The empirical \temp{O}{III} temperatures are found to have an average (0.95 $\pm$ 0.13) $\times$ $10^{4}$\.K. Furthermore, the average gas temperature \tempMD\ is calculated using all 202 \temp{N}{II} detections across the sample; \tempMD\ is found to have an average of (0.63 $\pm$ 0.30) $\times$ $10^{4}$\,K. \citetalias{Delgado2023a} finds that the substantial difference between \temp{O}{III} and \tempMD\ may be due to the presence of temperature inhomogeneities. This may explain the $\sim\!3 \times 10^{3}$\,K difference between the empirically derived \temp{O}{III} and \tempMD\ in this study.

\subsubsection{\texorpdfstring{$\rm [S\, III]\lambda$}{} \hspace{-2mm} 6312 Quality Check}
\label{subsec:siii_te_quality_check}

As mentioned in Section \ref{subsec:auroral}, we test the quality of $\rm [S\, III]\lambda$6312 auroral line detections by comparing the \temp{S}{III} -- \temp{N}{II} relationship with those found in the literature. The relation \temp{S}{III} -- \temp{N}{II} in Equation 8 from \cite{Rogers2022} is used to test the quality of the \temp{S}{III} measurements in this work; this is done under the assumption that \temp{N}{II} is relatively unaffected by telluric contamination. We find that on average, our \temp{S}{III} is only 123 $\pm$ 93 K higher than what is predicted by the relation by \cite{Rogers2022}. This relatively small difference between our direct \temp{S}{III} and the empirical \temp{S}{III} from Equation 8 by \cite{Rogers2022} demonstrates that our auroral line fitting method for $\rm [S\, III]\lambda$6312 produced reliable temperatures. 

\section{Ionic and Total Abundances}\label{sec:abundances}

In this section, the \dene and \te derived in Section \ref{sec:Electron Densities and Temperatures} are used to calculate \revtwo{various estimates} oxygen and nitrogen abundances. 

\subsection{Singly Ionized Oxygen Abundance Derivation}\label{subsec:OII_abundances}

\subsubsection{O\texorpdfstring{$^{+}$}{O+}Methodology}
\label{subsec:OII_methods}

We use various combinations of \temp{N}{II}, \temp{O}{II}, \denoii\ and \den{S}{II} along with emission line fluxes to directly derive four \revtwo{different estimates of the} O$^{+}$ abundances:

\begin{enumerate}
\item \temp{N}{II}, \denoii, $\rm [O\, II]\lambda$3727
\item \temp{O}{II}, \den{S}{II}, $\rm [O\, II]\lambda$3727
\item \temp{N}{II}, \den{S}{II}, $\rm [O\, II]\lambda$3727
\item \temp{N}{II}, \den{S}{II}, $\rm [O\, II]\lambda$7325
\end{enumerate}

\noindent Each \revtwo{estimate of the} O$^{+}$ abundance is derived using the \texttt{getIonAbundance} method from PyNeb. We propagate uncertainties for each of the quantities using the MC uncertainty method described in Section \ref{subsec:temp_den_methods}. 

\begin{figure*}
\centering
  \includegraphics[width=0.75\textwidth]{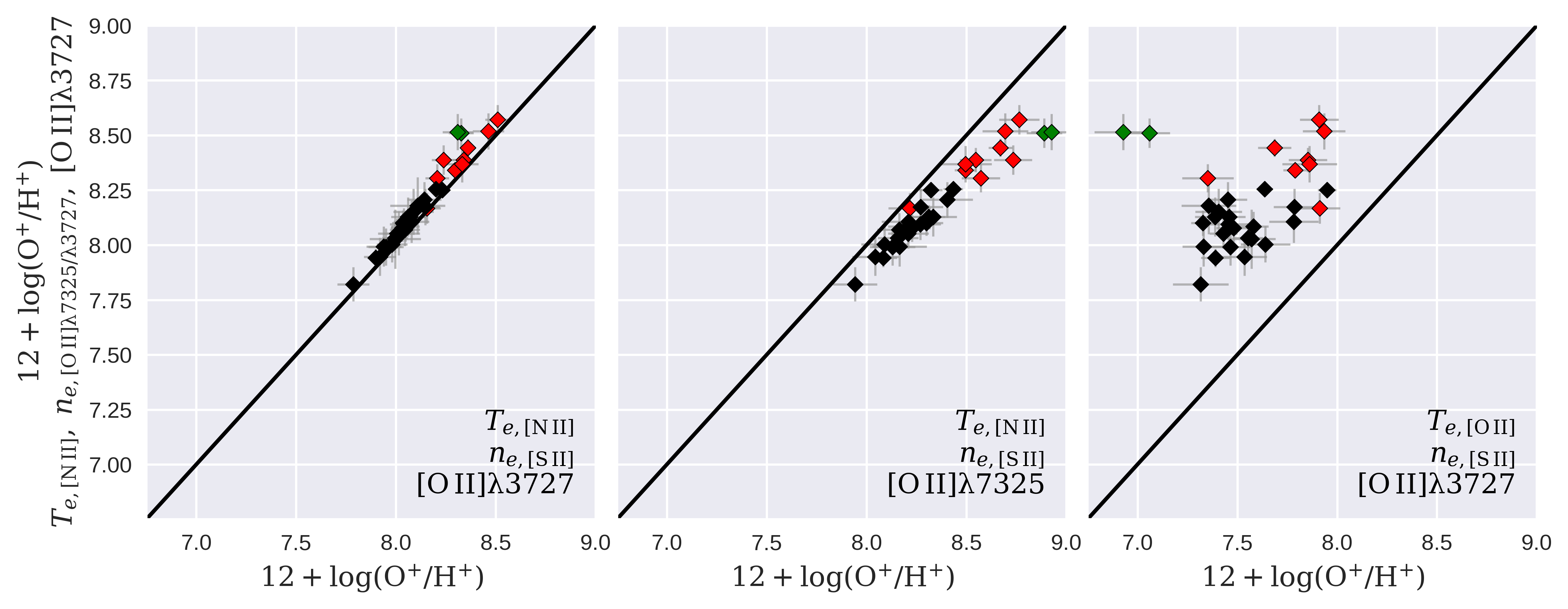}
  \caption{The reference O$^{+}$ abundance derived using \temp{N}{II}, $\rm [O\, II]\lambda$3727 and \denoii\ is shown on the y-axis plotted against our three other O$^{+}$ abundances described in Section \ref{subsec:OII_abundances}. The electron temperature, density, and emission lines used for each O$^{+}$ on the x-axis are printed in the bottom right corner of each panel. NGC~628 O$^{+}$ abundances are shown in red, NGC~2835 in black, and NGC4535 in green. There are no detections for NGC~3351 and NGC~3627. A 1-1 line is plotted in black to aid in visual comparison.}
  \label{fig:OII_abundance}
\end{figure*}

To create a basis of comparison for our O$^{+}$ derivations, we decide upon a `reference' O$^{+}$ abundance. 
\revtwo{First, because $\rm [O\,II]\lambda$7325 is exponentially sensitive to the electron temperature \citep{Osterbrock}, we adopt $\rm [O\,II]\lambda$3727 for the reference O$^{+}$ abundance derivation. This choice minimizes the dependence of the inferred O$^{+}$ abundances on possible density/temperature inhomogeneiteis within the gas, as well as on observational uncertainties.} Additionally, we choose to use \temp{N}{II} for our reference O$^{+}$ derivation rather than \temp{O}{II} to, once again, minimize the effect of \revtwo{any potential} density/temperature inhomogeneities \citepalias{Delgado2023b} on the O$^{+}$ abundance. Finally, we decide to use \denoii\ as the density in our reference O$^{+}$ instead of the widely used \den{S}{II}. This choice is motivated by the findings of \citetalias{Delgado2023b} and \citetalias{Vaught2023}, which \revtwo{suggest} that \den{S}{II} has a tendency to underestimate the average \dene, which can cause a $\sim$0.1 dex underestimation of the total oxygen abundance \citepalias{Delgado2023b}. We present our O$^{+}$ abundances in Figure \ref{fig:OII_abundance}, with our reference O$^{+}$ plotted against each of the other four O$^{+}$ abundance estimates. It is noted that the O$^{+}$ calculation that makes use of \temp{N}{II}, \den{S}{II} and $\rm [O\, II]\lambda$7325 is the only O$^{+}$ that uses only emission lines from MUSE.

\subsubsection{\texorpdfstring{O$^{+}$}{O+}Results}
\label{subsec:O+_results}

In Figure \ref{fig:OII_abundance} we compare four distinct direct O$^{+}$/H$^{+}$ derivations using various combinations of \te, \dene, and $\rm [O\, II]$ line fluxes across 5,658 \hii\ region candidates in five galaxies. We use our reference O$^{+}$/H$^{+}$ (\temp{N}{II}, \denoii, $\rm [O\, II]\lambda$3727) as a means for comparison, as described in Section \ref{subsec:OII_abundances}. Since we have no detections of $\rm [O\, II]\lambda$7325 in NGC~3351 or NGC~3627, we have no detections of our reference O$^{+}$/H$^{+}$ in these galaxies. In total, there are 36 detections of the reference O$^{+}$/H$^{+}$ abundance. Of these 36 measurements, the average is 8.2 [12 + log(O$^{+}$/H$^{+}$)] with a standard deviation of 0.9 [12 + log(O$^{+}$/H$^{+}$)]. 

In the leftmost panel of Figure \ref{fig:OII_abundance}, the \dene used in the O$^{+}$/H$^{+}$ derivation is varied. The O$^{+}$/H$^{+}$ on the $x$-axis makes use of \den{S}{II}, which is found to have systematically lower electron densities than \denoii\ in Section \ref{subsec:den_results}. The density underestimate from \den{S}{II} is the cause of an average ~0.06 dex lower O$^{+}$/H$^{+}$ compared to our reference O$^{+}$/H$^{+}$. This result agrees well with the ~0.1 dex underestimate due to density inhomogenities found by \citetalias{Delgado2023b}.

In the center panel of Figure \ref{fig:OII_abundance}, the use of \den{S}{II} and $\rm [O\, II]\lambda$7325 flux is studied when determining O$^{+}$/H$^{+}$ abundance. This combination of physical quantities leads to a O$^{+}$/H$^{+}$ that is $\sim0.3$ dex larger than our reference O$^{+}$/H$^{+}$. This abundance on the $x$-axis makes use of emission lines that are only found in the MUSE spectral range, thus this results suggests that studies using MUSE-only emission lines may be overestimating O$^{+}$/H$^{+}$ abundances.

In the rightmost panel of Figure \ref{fig:OII_abundance}, we vary the temperature diagnostic used to derive the O$^{+}$/H$^{+}$ abundance by using \temp{O}{II}. Since \temp{N}{II} is used to derive \denoii, we must use \den{S}{II} in order to have a sensible O$^{+}$(\temp{O}{II}) abundance. As noted in Section \ref{subsec:temp_results}, the high $\rm [O\, II]$ temperatures we derive are likely due to density inhomogeneities, which leads to an average ~0.61 dex underestimate of O$^{+}$/H$^{+}$ when compared to our reference O$^{+}$/H$^{+}$ abundance. This underestimate is especially apparent in NGC~4535 with a $\sim$1.45 dex offset in a single \hii\ region. These results confirm the prediction of \citetalias{Delgado2023b} that \temp{O}{II} leads to an underestimate of O$^{+}$ abundance. \revtwo{Furthermore, the appreciable scatter seen in this relationship is due to the scatter in the \temp{N}{II}--\temp{O}{II} trend from Figure \ref{fig:temp_corner}.} Thus, we urge caution in the use of \temp{O}{II} and suggest using \temp{N}{II} instead. Furthermore, when the $\rm [N\, II]\lambda$5755/$\lambda$6584 and $\rm [O\, II]\lambda$7325/$\lambda$3727 diagnostics are available, it is suggested to use \temp{N}{II} and \denoii\ to derive O$^{+}$/H$^{+}$. 

\subsection{Doubly Ionized Oxygen Derivation}\label{subsec:OIII_abundances}

\subsubsection{\texorpdfstring{O$^{2+}$}{O2+}Methodology}
\label{subsec:OIII_results}

Four different O$^{2+}$ abundances are calculated using various combinations of the \dene and \te described in Sections \ref{subsec:temp_den_methods}--\ref{subsec:empirical_temperature}. All four of the O$^{2+}$ abundances are calculated using the $\rm [O\, III]\lambda$5007 emission line from the Nebular Catalog:

\begin{enumerate}
\item \tempMD, \denoii
\item \tempMD, \den{S}{II}
\item \temp{O}{III}, \denoii
\item \temp{O}{III}, \den{S}{II}
\end{enumerate}

\begin{figure*}
\centering
  \includegraphics[width=0.75\textwidth]{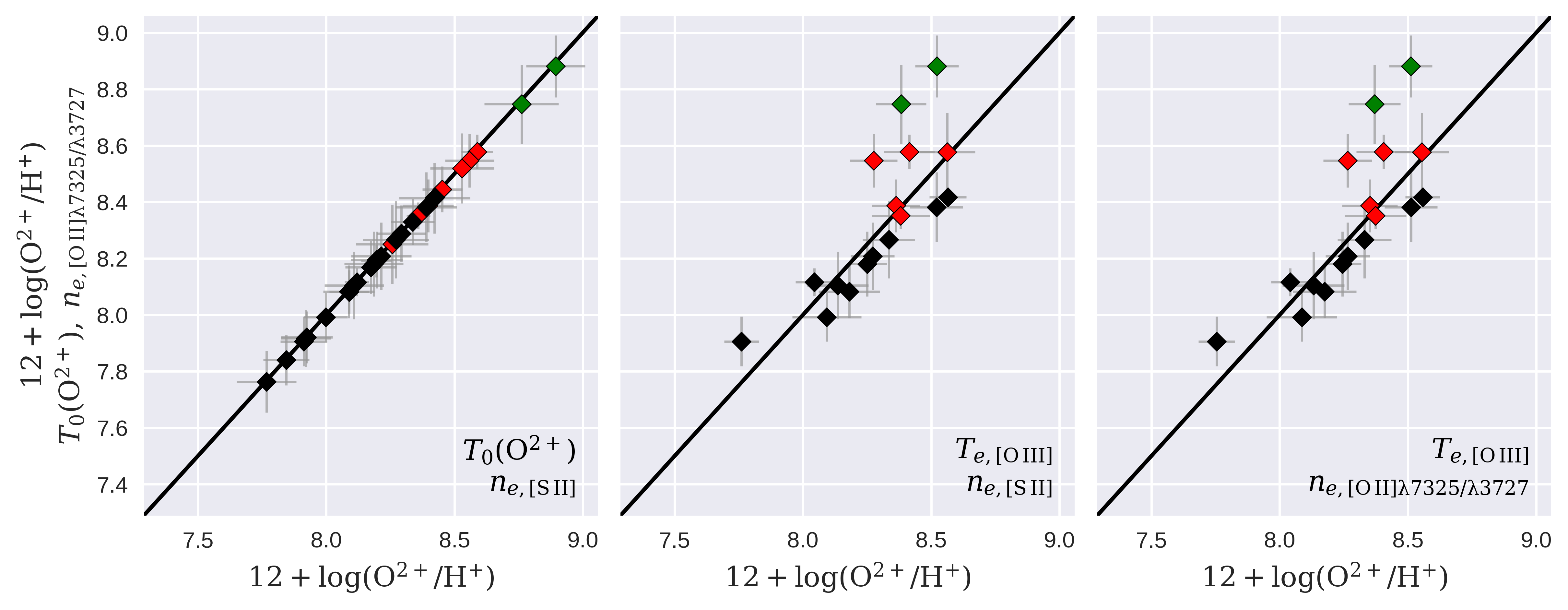}
  \caption{The reference O$^{2+}$ abundance derived using \tempMD\ and \denoii\ is shown on the y-axis plotted against our three other O$^{2+}$ abundances described in Section \ref{subsec:OIII_abundances}. The electron temperature, density, and emission lines used for each O$^{2+}$ on the x-axis are printed in the bottom right corner of each panel. NGC~628 O$^{2+}$ abundances are shown in red, NGC~2835 in black, and NGC4535 in green. There are no detections for NGC~3351 and NGC~3627. A 1-1 line is plotted in black to aid in visual comparison.}
  \label{fig:OIII_abundance}
\end{figure*}

The O$^{2+}$ that we choose to be our reference is the calculation that uses \tempMD\ and \denoii, as both of these quantities are expected to minimize the effect of temperature and density inhomogeneities, respectively (see Sections \ref{subsec:temp_den_methods} and \ref{subsec:empirical_temperature} for discussion). It should also be noted that all of the \hii\ regions in our sample with \temp{S}{III} detections have $\rm [O\, III] / [O\, II]$ $<$ 1.25. Figure 5 from \cite{Berg_2020} prioritizes empirical \temp{O}{III} in this regime as directly derived \temp{O}{III} show large scatter in \te -- \te for lower ionization regions. In Figure \ref{fig:OIII_abundance}, we plot our reference O$^{2+}$ against the three other O$^{2+}$ calculations. The O$^{2+}$ derivations that make use of \tempMD, \den{S}{II} and \temp{O}{III}, \den{S}{II} use emission lines that are only observed with MUSE. 

\subsubsection{\texorpdfstring{O$^{2+}$}{O2+}Results}
\label{subsec:O2+_results}

In Figure \ref{fig:OIII_abundance}, four different O$^{2+}$/H$^{+}$ derivations are compared. The O$^{2+}$/H$^{+}$ derived using \tempMD, \denoii\ and $\rm [O\, III]\lambda$5007 is taken to be our reference for doubly ionized oxygen abundance, as the first two quantities are expected to be the least affected by temperature and density inhomogeneities. \revtwo{It is found that our reference O$^{2+}$/H$^{+}$ estimates are in the range 7.7--8.9 [12 + log(O$^{2+}$/H$^{+}$)] and} an average of 8.30 [12 + log(O$^{2+}$/H$^{+}$)] among 30 \hii\ regions. Each of these 30 \hii\ regions also have a corresponding O$^{+}$/H$^{+}$ measurement, which has the same average and standard deviation from Section \ref{subsec:O+_results}. Thus, on average there is more O$^{2+}$/H$^{+}$ than O$^{+}$/H$^{+}$ in our sample of \hii\ regions. 

In the \revtwo{left} panel, our reference is compared to a O$^{2+}$/H$^{+}$ that is instead derived using \den{S}{II}. The trend shown in the \revtwo{left} panel of Figure \ref{fig:OIII_abundance} shows a nearly 1-1 correlation, indicating that the impact of density inhomogeneities on the resultant O$^{2+}$/H$^{+}$ is small, with only a $\sim$0.02 dex difference. \revtwo{This is expected because the critical density of the O$^{2+}$ $^1D_2$ level, which gives rise to $\rm [O\, III]\lambda5007$, is well above $10^3$ cm$^{-3}$.} This is promising for the PHANGS-MUSE program as O$^{2+}$/H$^{+}$ can be derived without significant impact from density or temperature inhomogeneities using \den{S}{II}, \tempMD\ , and $\rm [O\, III]\lambda$5007. 

On the contrary, the O$^{2+}$/H$^{+}$ derivations in \revtwo{the middle and right panels} of Figure \ref{fig:OIII_abundance} show deviations from our reference doubly ionized oxygen abundance. The O$^{2+}$/H$^{+}$ on the x-axis of \revtwo{the middle and right panels} both make use of the empirical \temp{O}{III} as well as \den{S}{II} and \denoii, respectively. \revtwo{These} O$^{2+}$/H$^{+}$ \revtwo{estimates} differ by $\sim$0.02 dex, \revtwo{ again because of the high critical density of the O$^{2+}$ $^1D_2$ level}. However, these derivations show that using \temp{O}{III} instead of \tempMD\ induces scatter in 12 + log(O$^{2+}$/H$^{+}$). In particular, this scatter increases for higher metallicity (12 + log(O$^{2+}$/H$^{+}$) \textgreater\ 8.3) \hii\ regions. Despite this scatter, the O$^{2+}$ derived from \temp{O}{III} and \tempMD\ agree quite strongly. This suggests that Equations \ref{eq:oiii_Te_B24} and \ref{eq:oiii_Te_MD23} can both be used to reliably estimate O$^{2+}$ for low ionization (i.e., $\rm [O\, III]/[O\, II]$ $<$ 1.25) \hii\ regions. 

\subsection{Elemental Oxygen Abundances}
\subsubsection{Methodology}\label{subsec:metal}

The total oxygen abundance is calculated \revtwo{using the ionization correction factor (ICF) module in PyNeb, specifically the \texttt{icf.getElementAbundance} method} \citep{pyneb}. An ICF from \cite{Izoto2006} is used to correct for unseen ionization levels. It is not expected that there will be meaningful contributions from O$^{3+}$ or higher oxygen ionization states, thus this ICF is the sum of O$^{+}$ and O$^{2+}$. We calculate the oxygen elemental abundance using each of the direct methods shown in Table \ref{tbl:total_abundances}. Where the first total abundance in Table \ref{tbl:total_abundances} uses the reference ionic abundances described in Sections \ref{subsec:OII_abundances} and \ref{subsec:OIII_abundances} while the third total abundance in Table \ref{tbl:total_abundances} uses MUSE-only derived quantities. MC uncertainties are determined by propagating the ionic abundance MC uncertainties. 

\begin{figure*}
\centering
  \includegraphics[width=\textwidth]{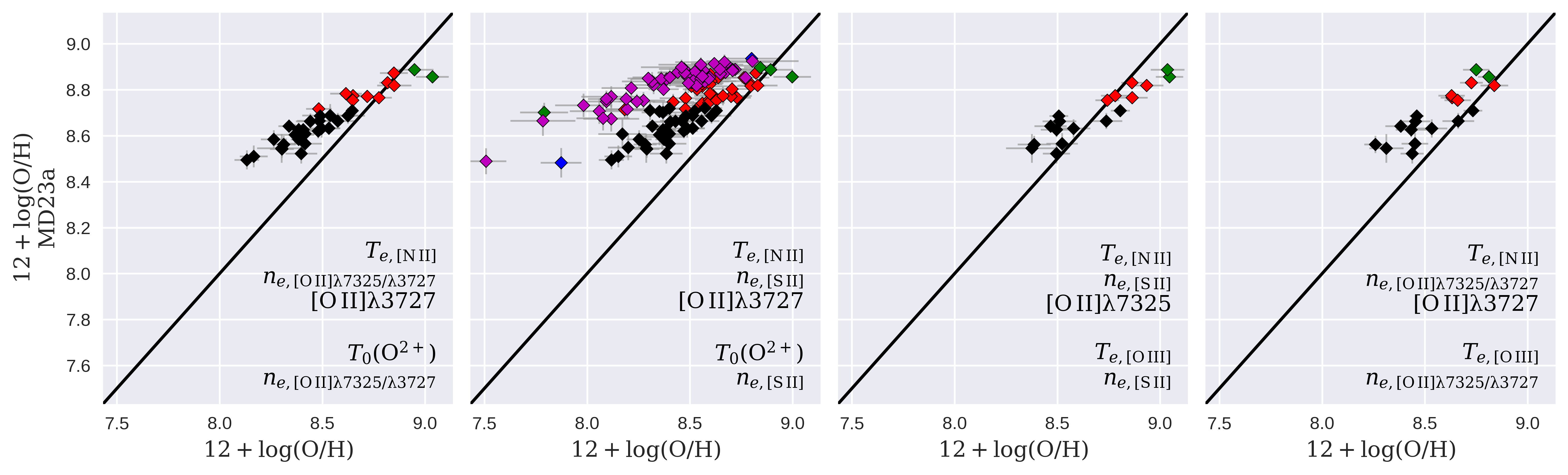}
  \caption{The \citetalias{Delgado2023a} metallicity calibration is on the y-axis is plotted against CEL direct metallicities. The electron temperature, density, and emission lines used for each metallicity on the x-axis is printed in the bottom right corner of each panel. The \citetalias{Delgado2023a} calibration plotted against our CEL direct abundances gives a proxy for the discrepancy between RL and CEL derived abundances. NGC~628 metallicites are shown in red, NGC~2835 in black, NGC4535 in green, NGC~3351 in blue, and NGC~3627 in purple. 1-1 lines are plotted in black to aid in visual comparison.}
  \label{fig:Direct_OH}
\end{figure*}

In \citetalias{Delgado2023a}, RL- and CEL-based metallicities are derived from high-S/N spectra of extragalactic and Galactic \hii\ regions. From the extragalactic \hii\ regions alone, \citetalias{Delgado2023a} derive an empirical calibration between \temp{N}{II} and RL-based metallicity. We adopt this calibration as a reference for our direct metallicity measurements because it is tied to RL-derived abundances. Comparison with the \citetalias{Delgado2023a} relation therefore serves as a proxy for assessing which direct CEL metallicities are most discrepant from RL-based values. Figure \ref{fig:Direct_OH} shows this comparison for all direct metallicity measurements with S/N $> 3$.

\begin{table*}
\caption{Elemental Oxygen abundance prescriptions considered in this work.}
\footnotesize
\centering
%\hspace*{-0.08\textwidth}
\begin{tabular}{@{}c c c c @{\hspace{0.8cm}} c c@{}}
\toprule
& \multicolumn{3}{c}{O$^{+}$} & \multicolumn{2}{c}{O$^{2+}$} \\
\cmidrule(r){2-4} \cmidrule(l){5-6}
 & $T_{e}$ & $n_{e}$ & Emission Line & $T_{e}$ & $n_{e}$ \\
\midrule
1 & \temp{N}{II} & \denoii\ & $\rm [O\, II]\lambda$3727 & \tempMD\ & \denoii\ \\
2 & \temp{N}{II} & \den{S}{II} & $\rm [O\, II]\lambda$3727 & \tempMD\ & \den{S}{II} \\
3 & \temp{N}{II} & \den{S}{II} & $\rm [O\, II]\lambda$7325 & \temp{O}{III} & \den{S}{II} \\
4 & \temp{N}{II} & \denoii\ & [$\rm [O\, II]\lambda$3727 & \temp{O}{III} & \denoii\ \\
\bottomrule
\end{tabular}
\label{tbl:total_abundances}
\end{table*}

\subsubsection{Results}
\label{subsec:OH_results}

The four direct elemental oxygen abundances, or metallicities, that we derive in Section \ref{subsec:metal} are compared with the \citetalias{Delgado2023a} calibration in Figure \ref{fig:Direct_OH}. The \citetalias{Delgado2023a} calibration serves as a proxy for an RL derived abundance. The \citetalias{Delgado2023a} calibration is found to have a mean value of 8.7 [12 + log(O/H)] and a standard deviation of 0.32 [12 + log(O/H)]. It should be noted that the range in which the \citetalias{Delgado2023a} relation was calibrated used \temp{N}{II} in the range \revtwo{0.8 -- 1.3 $\times10^4$~K, while our range of \temp{N}{II} is 0.6 -- 1.3 $\times10^4$~K}. This indicates that our sample may include \hii\ regions that have lower electron temperatures (i.e., higher metallicities), which may not be represented by the \citetalias{Delgado2023a} relation. 

In the leftmost panel of Figure \ref{fig:Direct_OH}, we compare the direct metallicity derived from our reference O$^{+}$ (\temp{N}{II}, \denoii, $\rm [O\, II]\lambda$3727) and reference O$^{2+}$ (\tempMD, \denoii) abundances with the \citetalias{Delgado2023a} calibration. As discussed in Sections \ref{subsec:OII_abundances} and \ref{subsec:OIII_abundances}, this direct CEL-derived metallicity should be relatively insensitive to temperature and density inhomogeneities. However, this method is on average $\sim$0.2 dex lower than the \citetalias{Delgado2023a} calibration. Despite this systematic offset, these two methods are still strongly correlated with $\rho$ = 0.91 $\pm$ 0.05, likely due to both methods having a strong dependence on \temp{N}{II}. This discrepancy appears to be pronounced at metallicities less than 8.75 [12 + log(O/H)], while at metallicities larger than 8.75 [12 + log(O/H)] it is found that the direct method agrees well with the \citetalias{Delgado2023a} calibration. Because both approaches are designed to trace the same underlying oxygen abundance and should be relatively insensitive to temperature and density inhomogeneities, their residual offset is somewhat unexpected. A more rigorous study will require both CEL and RL diagnostics to assess whether CEL-derived metallicities in the 12 + log(O/H) $>$ 8.75 regime are actually larger than RL derived metallicities. 

In the second panel of Figure \ref{fig:Direct_OH}, the \citetalias{Delgado2023a} calibration is compared to direct oxygen abundances derived using O$^{+}$ (\temp{N}{II}, \den{S}{II}, $\rm [O\, II]\lambda$3727) and reference O$^{2+}$ (\tempMD, \den{S}{II}). This is identical to the comparison in the first panel, except that \den{S}{II} is used in place of \denoii. For NGC~628, NGC~2835, and NGC~4535, the same overall trends are seen in both panels. Among the 35 \hii\ regions with direct metallicity estimates in both panels, the \den{S}{II}-based abundances are on average only $\sim$0.02 dex higher, indicating that replacing \denoii\ with \den{S}{II} produces only a minor systematic shift in the direct metallicities. Even so, this small offset remains systematic relative to the \citetalias{Delgado2023a} calibration. In addition, using \den{S}{II} increases the sample by 86 abundance measurements, enabling the inclusion of NGC~3351 and NGC~3627. The direct abundances in NGC~3627 are systematically lower than those of the rest of the sample relative to \citetalias{Delgado2023a}. This appears to be driven by its lower \temp{N}{II}: the median value for NGC~3627 is $\sim 7 \times 10^{4}$ K, compared to $\sim 8 \times 10^{4}$ K for NGC~628, NGC~2835, and NGC~4535. Although the origin of this $\sim 10^{4}$ K difference is unclear, the lower \temp{N}{II} values in NGC~3627 appear to drive the discrepancy between its direct abundances and the \citetalias{Delgado2023a} relation.

In the third panel of Figure \ref{fig:Direct_OH}, direct metallicities are derived using only emission lines covered by MUSE. This MUSE-only method adopts O$^{+}$ (\temp{N}{II}, \den{S}{II}, $\rm [O\, II]\lambda$7325) and O$^{2+}$ (\temp{O}{III}, \den{S}{II}). Relative to the reference metallicities in the first panel of Figure \ref{fig:Direct_OH}, the MUSE-only abundances are on average $\sim$0.01 dex higher. This close agreement is driven by compensating differences in the ionic abundances: $\rm [O\, II]\lambda$7325 produces a higher O$^{+}$ abundance than $\rm [O\, II]\lambda$3727, while \temp{O}{III} produces a lower O$^{2+}$ abundance relative to \tempMD. These opposite shifts largely cancel, so the total oxygen abundance remains similar to the reference value. Furthermore, the MUSE-only method agrees well with the \citetalias{Delgado2023a} calibration, with an average difference of $\sim$0.02 dex. This strong agreement is surprising given the impact of density inhomogeneities on \den{S}{II}, but indicates that the PHANGS-MUSE elemental oxygen abundances derived by \citetalias{Brazzini_2024} are not significantly affected by inhomogeneities in the gas. 

In the fourth panel of Figure \ref{fig:Direct_OH}, oxygen abundances are derived using O$^{+}$ (\temp{N}{II}, \den{S}{II}, $\rm [O\, II]\lambda$3727) and O$^{2+}$ (\temp{O}{III}, \den{S}{II}). This method differs from that used in the first panel only by adopting \temp{O}{III}, rather than \tempMD, for the O$^{2+}$ abundance calculation. The resulting \temp{O}{III}-based metallicities remain in strong agreement with the \tempMD-based metallicities. This method results in a $\sim$0.13 dex offset from the \citetalias{Delgado2023a} calibration. This result shows that \temp{O}{III} or \tempMD\ yield similar metallicities within our sample and that the ionization-based temperature priorities prescribed by \cite{Berg_2020} should be used. Furthermore, it seems that using either \temp{S}{III} and Equation \ref{eq:oiii_Te_B24} or \temp{N}{II} and Equation \ref{eq:oiii_Te_MD23} yield not only reliable O$^{2+}$, but also total metallicities. 

Each of the metallicity prescriptions in Table \ref{tbl:total_abundances}, except for the second metallicity, yield oxygen abundances that agree within a few tenths of a dex relative to \citetalias{Delgado2023a}. This indicates that these metallicities are not significantly affected by temperature or density inhomogeneities. However, the \revtwo{leftmost} and \revtwo{rightmost} metallicities in Figure \ref{fig:Direct_OH} show linear and constant systematic offsets, respectively, in comparison to \citetalias{Delgado2023a}. \revtwo{One potential mechanism that could contribute to these systematic offsets is DIG contamination 
(see Section~\ref{subsec:den_results}). However, a careful analysis of metallicity measurements 
with and without DIG subtraction is required to quantify its contribution.} While the MUSE-only metallicity method exhibits agreement with \citetalias{Delgado2023a}, but this appears to be coincidental rather than physically motivated. Specifically, \den{S}{II} and $\rm [O\, II]\lambda$7325 produce a larger O$^{+}$ abundance than the reference method for more enriched regions, while \temp{O}{III} yields a lower O$^{2+}$ abundance relative to \tempMD\ for these same regions. The underlying cause for the differences between the direct methods and \citetalias{Delgado2023a} in Figure \ref{fig:Direct_OH} must be explored more closely for datasets containing spectra which include both RLs and CELs that can be used for temperature and density calculations. 

\subsection{Nitrogen Abundances}

\subsubsection{Methodology}
\label{subsec:nitrogen}

Alongside the various oxygen abundances, we directly calculate the singly ionized nitrogen abundances $\rm N^{+}$. This calculation uses \temp{N}{II} and the strong nebular $\rm [N\, II]\lambda$6584 line. To evaluate the effect of density variations on $\rm N^{+}$, we perform two separate calculations using \den{S}{II} and \denoii. The $\rm N^{+}$ abundance is derived from the \texttt{getIonAbundance} method in PyNeb. We calculate \denoii\ derived $\rm N^{+}$/O$^{+}$ abundance ratio using O$^{+}$(\temp{N}{II}, \denoii,  $\rm [O\, II]\lambda$3727). Similarly, we use O$^{+}$(\temp{N}{II}, \den{S}{II},  $\rm [O\, II]\lambda$3727) for the \den{S}{II} derived $\rm N^{+}$/O$^{+}$. 

The elemental nitrogen abundance, $\rm N$, is obtained using an ICF calibrated in \cite{Amayo2021}. The ICF is defined as: 

\begin{gather}
\label{eq:nitrogen_icf_poly}
\begin{aligned}
\log\left(\rm ICF\left(N^{+}/O^{+}\right)\right) = & \text{  }0.013 - 0.793\omega\\ &+ 8.177\omega^{2}
- 23.194\omega^{3}\\ &+ 26.364\omega^{4} - 10.536\omega^{5},
\end{aligned}
\end{gather}

\noindent where $\omega$ = O$^{2+}$ / (O$^{+}$ + O$^{2+}$). We find the elemental abundance with:

\begin{gather}
\begin{aligned}
\label{eq:nitrogen_icf}
\rm \cfrac{N}{O} = \cfrac{N^{+}}{O^{+}} \times ICF\left(N^{+}/O^{+}\right).
\end{aligned}\raisetag{0\baselineskip}
\end{gather}

\begin{figure*}
\centering
  \includegraphics[width=0.7\textwidth]{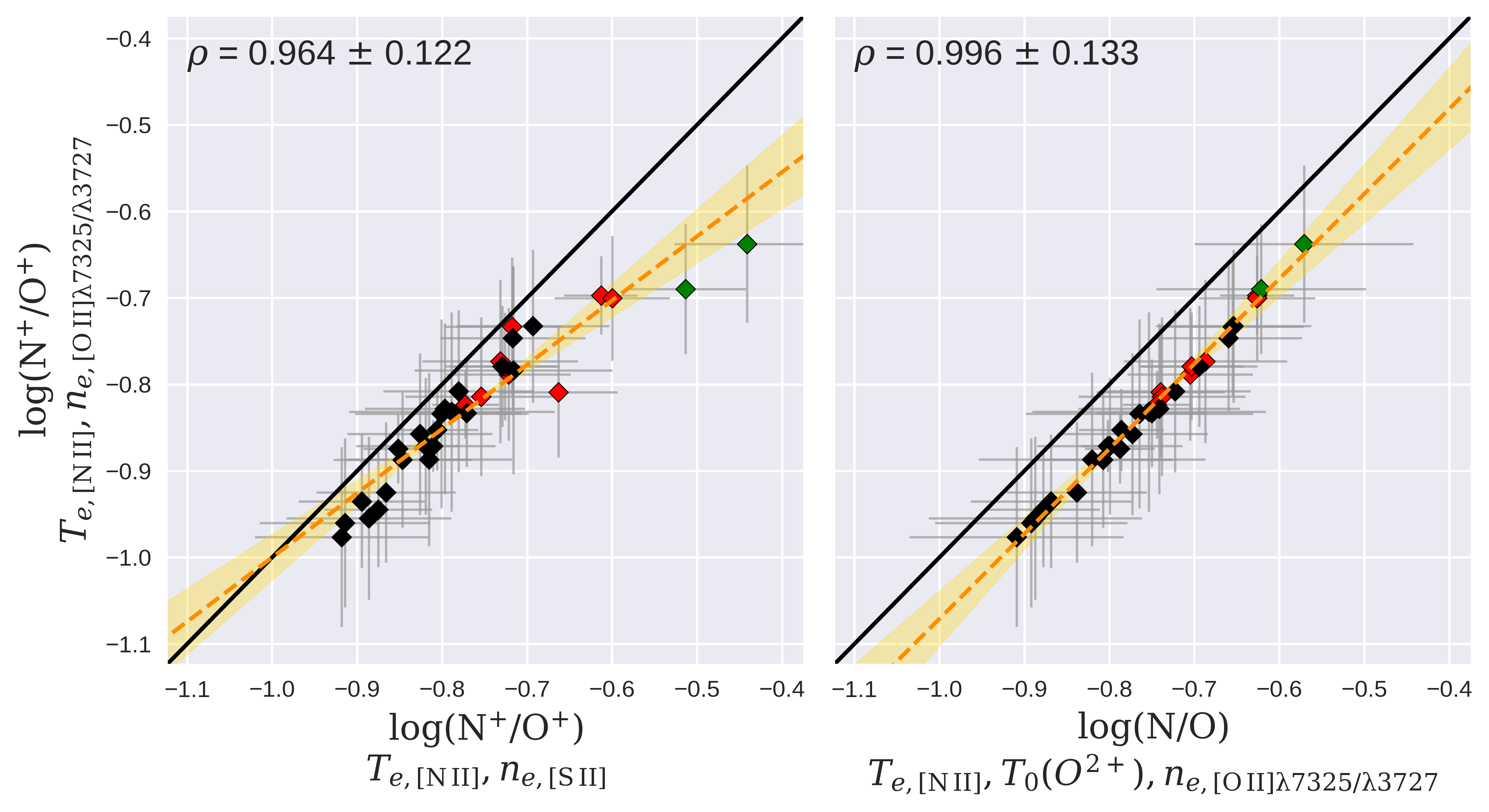}
  \caption{On the y-axis $\rm N^{+} / O^{+}$ (\denoii) is plotted against $\rm N^{+} / O^{+}$ (\den{S}{II}) in the left panel and the elemental nitrogen abundance $\rm N / O$ in the right panel. NGC~628 \hii\ region data points are shown in red, NGC~2835 in black, and NGC4535 in green; no  NGC~3351 or NGC~3627 \hii\ regions had detections. A linear polynomial is fit to each relation, shown by the gold dotted line, with the 1 $\sigma$ uncertainty in the fits shaded in gold. The Spearman rank correlation coefficient and the respective uncertainties are shown in the upper left corner of each plot. The 1-1 line is shown in black.}
  \label{fig:nitrogen_comp}
\end{figure*}

\begin{figure*}
\centering
  \includegraphics[width=0.75\textwidth]{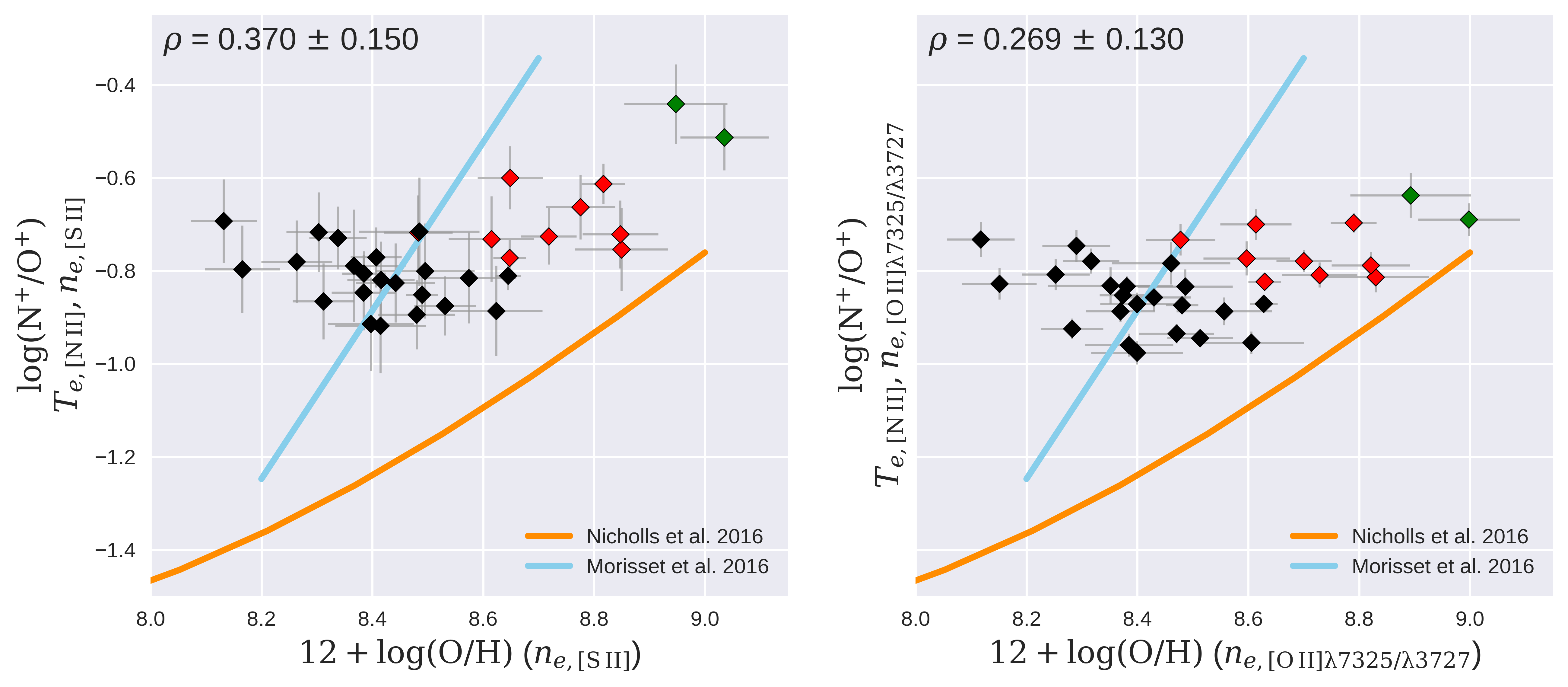}
  \caption{In the left panel, $\rm N^{+} / O^{+}$ (\den{S}{II}) is plotted against the second direct metallicity in Table \ref{tbl:total_abundances}; the right panel plots $\rm N^{+} / O^{+}$ (\den{S}{II}) against the first direct metallicity prescription Table \ref{tbl:total_abundances}. NGC~628 \hii\ region data points are shown in red, NGC~2835 in black, and NGC~4535 in green. There are no detections of NGC~3351 and NGC~3627. The predictions for N/O -- O/H relations from \cite{Nicholls2016} and \cite{Morisset2016} are plotted in orange and cyan, respectively.}
  \label{fig:nitrogen_plots}
\end{figure*}

\noindent We use O$^{+}$(\denoii, \temp{N}{II}, $\rm [O\, II]\lambda$3727) and O$^{2+}$ (\denoii, \tempMD) to derive $\omega$ and calculate N/O. For all nitrogen abundances described in this section we propagate uncertainties using the same MC methods described in Section \ref{subsec:temp_den_methods}. We compare each of the Nitrogen abundances discussed in this section in Figure \ref{fig:nitrogen_comp} and plot different N$^{+}$/O$^{+}$ prescriptions against their corresponding metallicity in Figure \ref{fig:nitrogen_plots}. 

\subsubsection{Results}
\label{subsec:N_results}

In Figure \ref{fig:nitrogen_comp}, the log(N$^{+}$/O$^{+}$) ratio derived using \denoii\ and \temp{N}{II} is shown on the y-axis and is used as the \revtwo{reference to minimize potential biases from density inhomogeneities}. In regions with log(N$^{+}$/O$^{+}$) abundance detections derived from \denoii\ and \den{S}{II}, the \den{S}{II}-derived log(N$^{+}$/O$^{+}$) is found to be, on average, $\sim$0.06 dex higher. In the left panel of Figure \ref{fig:nitrogen_comp} we can see a direct comparison of these methods. The only difference in these derivations is the use of the density diagnostic for both N$^{+}$ and O$^{+}$. Thus, we can attribute this $\sim$0.06 dex difference in log(N$^{+}$/O$^{+}$) methods to potential density inhomogeneities.

In the right panel of Figure \ref{fig:nitrogen_comp}, the N$^{+}$/O$^{+}$(\denoii) is compared to the total elemental log(N/O) derived by applying the ICF from Equation \ref{eq:nitrogen_icf_poly} and \ref{eq:nitrogen_icf}. In the 34 regions in which there are measurements of both N$^{+}$/O$^{+}$(\denoii) and log(N/O), averages of -0.82 $\pm$ 0.08 dex and -0.75 $\pm$ 0.08 dex are found for the respective ratios. This offset of $\sim$0.08 dex is nearly constant in these 34 \hii\ regions. \revtwo{This suggests that, for our sample, the variation due to density prescription in log(N$^{+}$/O$^{+}$) is comparable in magnitude to the variation introduced by the ICF used to infer log(N/O). Therefore, caution should be taken in what density prescription is used when deriving log(N$^{+}$/O$^{+}$). Additionally,} it is a common assumption \citep{Peimbert1967, Thurston1996} to set log(N/O) = log(N$^{+}$/O$^{+}$). However, this offset of $\sim$0.08 dex indicates the assumption that log(N/O) = log(N$^{+}$/O$^{+}$) may be a poor postulate. The ratio N$^{+}$/O$^{+}$(\denoii) / (N/O) is found to be 0.84 $\pm$ 0.01 in linear units. \revtwo{This differs from the photoionization model predictions shown in Figure~2 of \cite{Thurston1996}, where the ratio of N$^{+}$/O$^{+}$ to N/O remains greater than 0.95 for the model conditions explored in that work. The exact value of N$^{+}$/O$^{+}$ to N/O is model-dependent, however our results illustrate that the assumption log(N/O) = log(N$^{+}$/O$^{+}$) should be applied with care.}

%When comparing N$^{+}$/O$^{+}$ derived with \den{S}{II} to the elemental N/O, a ratio of 0.97 $\pm$ 0.11 is found. This result is in much better agreement with \cite{Thurston1996}. This suggests that N$^{+}$/O$^{+}$(\den{S}{II}) $\approx$ (N/O). This agreement can be attributed to increased N$^{+}$/O$^{+}$(\den{S}{II}) due to \den{S}{II} being underestimated relative to \denoii (see Section \ref{subsec:den_results}). 

%The ratio obtained from N$^{+}$/O$^{+}$(\denoii) / (N/O) implies that there is a non negligible amount of either neutral or doubly ionized nitrogen in our sample of \hii\ regions. Given the ionized nature of \hii\ regions, we speculate that this additional nitrogen abundance is most likely of the doubly ionized species. A more in-depth study on where this additional nitrogen abundance must be done in order to understand the difference between N$^{+}$/O$^{+}$(\denoii) and N/O. 

In Figure \ref{fig:nitrogen_plots} we study the relation between N/O and O/H. In general, nitrogen is produced primarily through Type II supernova, but has a secondary production mechanism through the evolution of intermediate-mass stars. When the abundance of nitrogen increases faster than that of oxygen, this signifies a chemically evolved ISM due to additional nitrogen production from the evolution of intermediate mass stars. Equation 3 from \citet[][hereafter N16]{Nicholls2016} and Equation 4 from \citet[][hereafter M16]{Morisset2016} both describe this expected relationship between nitrogen and oxygen as a function of oxygen abundance; both relations are overlaid in the right panel of Figure \ref{fig:nitrogen_plots}. The \citetalias{Nicholls2016} relation covers a much larger metallicity range of $\sim$6 $<$ 12 + log(O/H) $<$ $\sim$9, while \citetalias{Morisset2016} covers the metallicity range 8.1 $<$ 12 + log(O/H) $<$ 8.8. The \citetalias{Nicholls2016} relation is nonlinear, covering both ISMs with nitrogen contribution from only Type II supernovae and chemically evolved ISMs, contrarily the \citetalias{Morisset2016} relation only encompasses the metallicity range including chemically evolved ISMs. Both the \citetalias{Nicholls2016} and \citetalias{Morisset2016} relations are derived using outputs from photoionization models that are constrained using observations; the \citetalias{Nicholls2016} relation is constrained using MW stellar nebular data and \citetalias{Morisset2016} is constrained using galaxies in the local universe. 

All of our \hii\ regions in Figure \ref{fig:nitrogen_plots} have metallicty 12 + log(O/H) $>$ 8.1, such that both \citetalias{Morisset2016} and \citetalias{Nicholls2016} imply that nucleosynthesis from evolved intermediate-mass stars are contributing to the nitrogen enrichment of the ISM. Figure \ref{fig:nitrogen_plots} shows the N/O--O/H relation derived using \den{S}{II} in the left panel and \denoii\ in the right panel. In both panels, all detections lie on the upper-left side of the \citetalias{Nicholls2016} curve, whereas the majority of points lie on the lower-right side of the \citetalias{Morisset2016} relation. \revtwo{This demonstrates that the majority of the \hii\ regions in our sample, with the exception of a handful of NGC~2835 regions, fall between the \citetalias{Nicholls2016} and \citetalias{Morisset2016} relations. This result is unsurprising given the large scatter found in the N/O -- O/H relation at high metallicities throughout the literature (e.g., \citealt{Berg_2020}).} 

In total, 35 \hii\ regions have N/O--O/H detections in both panels. Among these detections, we find that, on average, N$^{+}$/O$^{+}$(\den{S}{II}) and N$^{+}$/O$^{+}$(\denoii) are $\sim$0.39 dex and $\sim$0.31 dex higher than the \citetalias{Nicholls2016} relation, respectively, showing that both derivations differ substantially from \citetalias{Nicholls2016}. In comparison with \citetalias{Morisset2016}, N$^{+}$/O$^{+}$(\den{S}{II}) is lower by $\sim$0.06 dex, while N$^{+}$/O$^{+}$(\denoii) is lower by $\sim$0.17 dex. \revtwo{Although the average value agrees more closely with \citetalias{Morisset2016}, our sample appears to follow a functional form similar to that of \citetalias{Nicholls2016}. The offset between the relations may reflect differences in abundance methodology, for example the adopted ICF prescription, but a detailed assessment is beyond the scope of this work.}

The varying levels of agreement with \citetalias{Morisset2016} and \citetalias{Nicholls2016} are likely driven by differences in the samples used to construct the two relations. \citetalias{Morisset2016} derived its relation from an \hii\ region sample, similar to the one shown in Figure \ref{fig:nitrogen_plots}, whereas \citetalias{Nicholls2016} combined stellar and \hii\ region detections. \revtwo{\citetalias{Brazzini_2024} found good agreement with \citetalias{Nicholls2016} using PHANGS-MUSE data. This agreement may reflect differences in the abundance methodology, since \citetalias{Brazzini_2024} adopt an ICF prescription from \cite{Izoto2006}, whereas we study 
N$^{+}$/O$^{+}$ directly. In contrast, the relation by \cite{Berg_2020} demonstrates} more scatter in the NO--OH relation. In particular, regions from NGC~628 in our sample and \cite{Berg_2020} seem to occupy the same place in NO-OH space. \revtwo{Although methodological differences may play a role, the large intrinsic scatter in the N/O--O/H relation likely also contributes to the differences between our sample and the aforementioned works.}

\section{Discussion}
\label{sec:discussion}

In this section, we discuss our findings in the context of previous studies, with an emphasis on discussing how temperature and density inhomogeneities contribute to the ADF.

\subsection{Galaxy Parameter Splitting}

The galaxies across our comparisons of electron densities, electron temperatures, and abundances consistently populate different regions of parameter space. For example, in the case of \denoii, NGC~2835 has an average value of $700 \pm 200 \text{ cm}^{-3}$, NGC~628 has an average value of $1100 \pm 600 \text{ cm}^{-3}$, and NGC~4535 has an average value of $3200 \pm 200 \text{ cm}^{-3}$. A similar pattern is seen in \temp{N}{II}, with NGC~2835 having an average of $9000 \pm 800$ K, NGC~628 $7700 \pm 900$ K, NGC~4535 $7700 \pm 1700$ K, and NGC~3627 $7600 \pm 1000$ K. Similar trends are seen in the literature (e.g., see \citealt{Berg_2020, Vaught2023}), where the \te -- \te trends and various abundance relations demonstrate similar galaxy segregation. 

For \temp{N}{II}, all galaxies except NGC~2835 have average values that agree within the standard deviations, whereas the \denoii\ averages remain clearly separated from one another across galaxies. This result follows for \te, the oxygen abundances, and nitrogen abundances. This result suggests that \hii\ regions within the same galaxy tend to exhibit similar physical conditions, while those conditions may vary substantially between galaxies, as seen for \temp{N}{II} and \denoii\ in our sample. However, because our sample includes only five galaxies and provides a limited number of detections for these quantities, a more comprehensive analysis of \hii\ region properties both within and among galaxies is needed to better understand this apparent galaxy-to-galaxy variation. The PHANGS-MUSE sample presents an ideal dataset for such an analysis. 

\subsection{Density Inhomogeneities}
\label{subsec:den_inhomogeneities}

In Section \ref{subsec:den_results}, we found a significant discrepancy of $\sim10^{3}$ cm$^{-3}$ between \den{S}{II} and \denoii. It is possible to attribute this difference to density inhomogeneities within \hii\ regions, which can be understood in terms of the differing critical densities of the relevant transitions. The $\rm [O\, II]\lambda3727$ and $\rm [S\, II]\lambda\lambda6716,6731$ nebular lines have critical densities of order $\sim10^{3}$ cm$^{-3}$ \citep{pyneb}, whereas the $\rm [O\, II]\lambda$7325 auroral line remain density sensitive up to $\sim10^{5}$ cm$^{-3}$. As the gas density approaches or exceeds the critical density of a given transition, collisional de-excitation reduces its emissivity. If \hii\ regions contain dense clumps with \dene $\sim 10^{3}$--$10^{5}$ cm$^{-3}$, then the $\rm [S\, II]$ nebular lines will preferentially trace the lower-density gas, while the $\rm [O\, II]$ auroral lines can still retain sensitivity to denser material. This picture is consistent with Figure \ref{fig:den-den}, where \den{S}{II} is systematically biased toward lower values, often near the low-density limit, and naturally explains why \den{S}{II} yields lower densities than \denoii. 

The suppression of the $\rm [S\, II]\lambda\lambda6716,6731$ and $\rm [O\, II]\lambda3727$ nebular emission by density inhomogeneities may have important consequences for the derivation of \temp{O}{II}. When the $\rm [O\, II]\lambda7325/\lambda3727$ ratio is combined with the underestimated \den{S}{II}, the inferred \temp{O}{II} becomes biased toward the hotter, lower-density gas within the \hii\ region. By contrast, the nebular $\rm [N\, II]$ and $\rm [S\, III]$ lines have critical densities of $\sim10^{5}$ cm$^{-3}$ and $\sim10^{6}$ cm$^{-3}$, respectively, such that \temp{N}{II} and \temp{S}{III} remain largely insensitive to density inhomogeneities over the density range relevant for ionized nebulae. This picture may naturally explain the trends seen in Figure \ref{fig:temp_corner}, where \temp{O}{II} $>$ \temp{N}{II} $\approx$ \temp{S}{III}. Although the $\rm [O\, II]\lambda7325/\lambda3727$ ratio being unreliable as a temperature diagnostic when used with \den{S}{II}, the $\rm [O\, II]$ auroral-to-nebular ratio remains a powerful density diagnostic \citepalias{Delgado2023b}. Following theoretical predictions for regions with homogeneous \dene and setting \temp{O}{II} = \temp{N}{II}, the resultant \denoii\ yields an average electron density that is unaffected by density inhomogeneities.

\revtwo{As discussed in Section \ref{subsec:den_results}, the DIG may impact the observed fluxes of lower ionization lines such as $\rm [N\, II]\lambda$6548 and $\rm [O\, II]\lambda$3727 \citep{Zhang2017}. While both density inhomogeneities and DIG contamination may contribute to the abundance offsets observed in this work (e.g., Figure~\ref{fig:OII_abundance}), the present analysis does not allow us to quantify their relative importance. Additionally, it remains unclear whether these offsets depend on galactic environment. A follow up study will examine the relative contributions of these effects in more detail.}

\begin{figure}
\centering
  \includegraphics[width=0.475\textwidth]{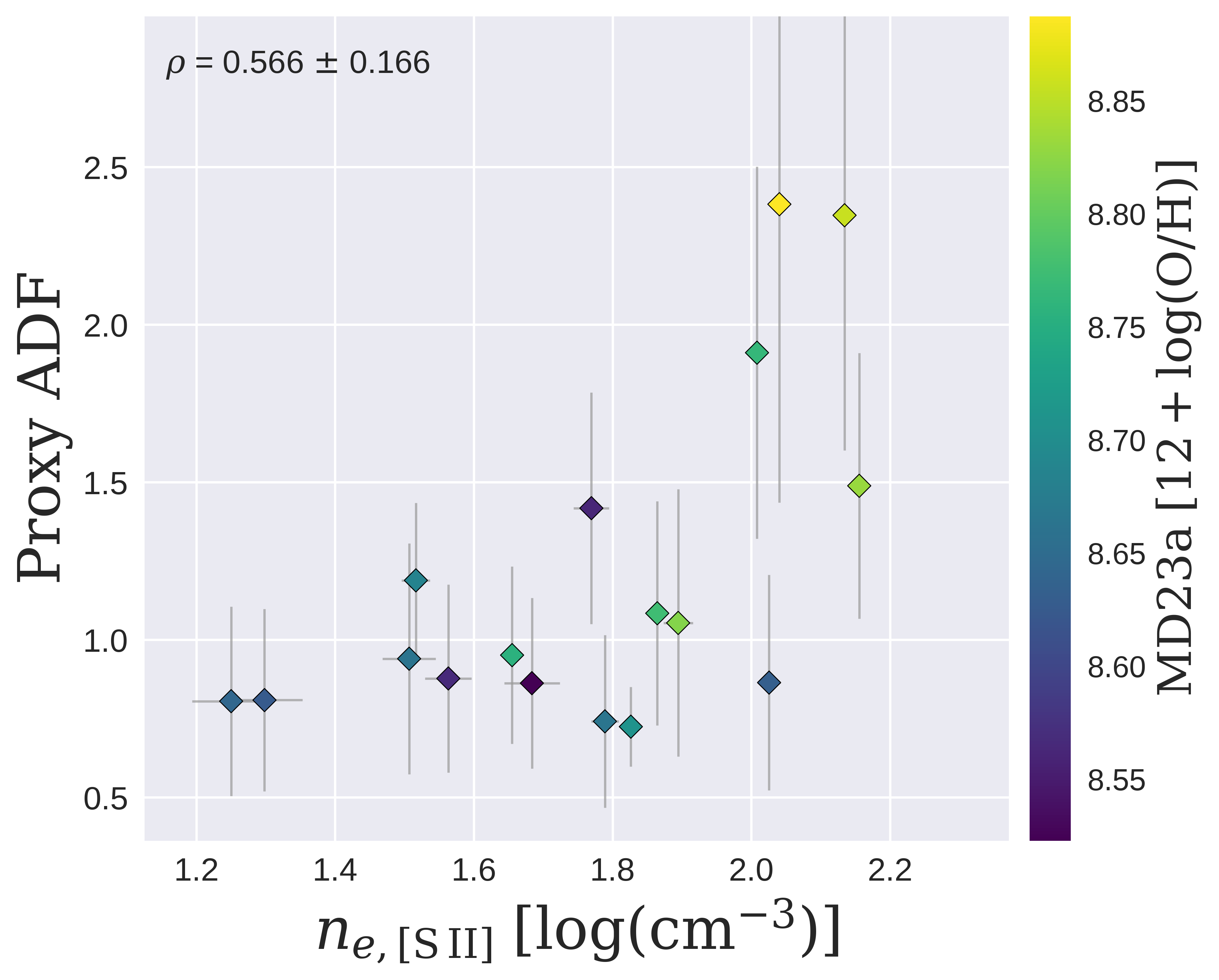}
  \caption{On the y-axis we plot the proxy ADF, defined as the ratio of O$^{2+}$ (\tempMD, \denoii) and O$^{2+}$ (\temp{O}{III}, \denoii). The proxy ADF is plotted against \den{S}{II} and colored by the \citetalias{Delgado2023a} metallicity calibration. The Spearman rank correlation coefficient in shown in the upper left corner of each plot.}
  \label{fig:adf_plot}
\end{figure}

\vspace{5mm}

\subsection{Temperature Inhomogeneities}
\label{subsec:temp_inhomogeneities}

The $\rm [O\, III]\lambda$4363 auroral line is outside the spectral range of MUSE and SITELLE, however the use of \temp{S}{III} and Equation \ref{eq:oiii_Te_B24} allow for \temp{O}{III} to be empirically calculated. Furthermore, this work does not make use of RLs, but uses the empirically derived \tempMD\ using \temp{N}{II} and Equation \ref{eq:oiii_Te_MD23}. In Figure \ref{fig:OIII_abundance}, we find that O$^{2+}$ derived from either \temp{O}{III} or \tempMD\ generally agree well with differences only up to $\sim0.05$ dex. Furthermore, the elemental oxygen abundances derived with O$^{2+}$ agree well with the \citetalias{Delgado2023a} metallicity calibration. However, we do find that there are systematic offsets between \citetalias{Delgado2023a} and the metallicities derived from \tempMD\ and \temp{O}{III} with these differences being more pronounced in lower metallicity regions following the results of \citetalias{Delgado2023a}. It is not clear what is the cause of the offsets seen in the first and fourth panels of Figure \ref{fig:Direct_OH}, however it seems that using \temp{O}{III} from \citetalias{Brazzini_2024} and \tempMD\ from \citetalias{Delgado2023a} minimize the impact of temperature inhomogeneities on metallicity. In future studies, a comparison between RL derived abundances and abundances derived using multiple photoionization models \citep{Marconi_2024} may reveal the cause of the discrepancies seen in Figure \ref{fig:Direct_OH}. 

\subsection{Proxy ADF}

To further study the differences in the derived O$^{2+}$ from \tempMD\ and \temp{O}{III}, a proxy for the ADF is defined as the ratio of O$^{2+}$ derived from \tempMD\ and \temp{O}{III}. We emphasize that this ratio is a proxy of the ADF as \tempMD\ and \temp{O}{III} are empirically derived temperatures. In Figure \ref{fig:adf_plot}, the proxy ADF is plotted against \den{S}{II}. \revtwo{The Spearman rank correlation coefficient is positive ($\rho_{\rm S} \sim 0.56$), suggesting a possible monotonic trend between the proxy ADF and \den{S}{II}. However, because this trend is based on only four noisy measurements, we interpret it with caution.} A similar trend between the proxy ADF and \denoii\ can be found, which indicates this \revtwo{possible} relationship is most likely not due to density inhomogeneities. If density inhomogeneities were one of the main contributors to this trend, then a relationship between the proxy ADF and \den{S}{II} might be expected, but not in \denoii\ . Howerver, if pockets of densities exist that exceed the $\rm [O\, II]$ auroral-to-nebular line critical density ($>10^5$ cm$^{-3}$), then density inhomogeneities may be contributing to the proxy ADF. 

Another possible explanation is the notion that H-deficient clumps with large densities may exist inside of metal rich \hii\ regions, i.e. chemical inhomogeneities. The existence of such clumps would bias observed RL metallicities to higher values, thus increasing an observed ADF. In Figure \ref{fig:adf_plot}, the regions are also colorized by their metallicity, which can also be seen to increase with the proxy ADF. Such a correlation suggests that the proxy ADF is more prevalent in chemically evolved ISMs. It is found that NGC~4535 has the largest proxy ADF, with values of $\sim$2.4, while NGC~628 has values of $\sim$1 and NGC~2835 has values of $\sim$1.5. The correlation between \dene and metallicity is suggestive that chemically enriched gas contributes to the proxy ADF. This does not imply the existence of H-deficient clumps, however this hypothesis can only be studied via spatially resolved observations of \hii\ regions, similar to the study carried out in \citetalias{Delgado2023a}, but for a sample of high metallicity (12 + log(O/H) $>$ 8.7) \hii\ regions. 

\section{Conclusion}
\label{sec:conclusion}

In this work, observations of five galaxies from SITELLE are paired with PHANGS-MUSE observations to study the properties of extragalacitic \hii\ regions. Various electron density, electron temperature and abundance determinations are compared within this work and to previous studies. The main findings in this work are as follows:

\begin{itemize}

    \item An $\rm [O\, II]$ density \denoii\ is derived using the $\rm [O\, II]$ auroral-to-nebular ratio in combination with the $\rm [N\, II]$ temperature. On average \denoii\ is found to be $\sim10^3$ cm$^{-3}$ larger than \den{S}{II}. This substantial underestimation in electron density from the $\rm [S\, II]$ diagnostic results in a $\sim0.1$ dex underestimation in O$^{+}$.

    \item We find that density inhomogeneities may \revtwo{underestimate} electron densities when using the $\rm [S\, II]$ nebular-to-nebular line diagnostic, which coincides with results from \citetalias{Delgado2023b} and \citetalias{Vaught2023}. When \den{S}{II} is paired with the $\rm [O\, II]$ auroral-to-nebular line ratio, then the $\rm [O\, II]$ temperature is overestimated by $\sim$0.3 $\times 10^4$~K due to density inhomogeneities. 

    \item The \revtwo{empirical temperatures} \tempMD\ \revtwo{from} \citetalias{Delgado2023a} and \temp{O}{III} \revtwo{from} \citetalias{Brazzini_2024} are found to produce O$^{2+}$ abundances that minimize the effects of temperature inhomogeneities.  

    \item The relation between N/O and N$^{+}$/O$^{+}$ is found to deviate modestly from the commonly adopted one-to-one approximation \revtwo{when \denoii is adopted as the density prescription}. For our sample, we find $\mathrm{N}^{+}/\mathrm{O}^{+} \sim (0.84 \pm 0.01)\mathrm{N}/\mathrm{O}$.

    \item A potential correlation is found between the proxy ADF and properties that describe chemical enrichment (i.e., metallicity and \dene). These results suggest that \hii\ regions that are more chemically evolved may have larger ADFs. This result needs to be rigorously tested with a direct calculation of O$^{2+}$ using both RL- and CEL-derived temperatures.
    
\end{itemize}

The findings of this work imply that temperature and density inhomogeneities remain a central issue in estimating chemical abundances in photoionized nebulae. \revtwo{However, this work demonstrates that the impact of these biases can be reduced by adopting temperature and density prescriptions that are less sensitive to temperature and density inhomogeneities. Any remaining biases appear to be limited to a few tenths of a dex, substantially smaller than the order of magnitude offsets that have historically affected some direct abundance determinations. Understanding the origin of these remaining biases and their connection to the} ADF remains a central goal of studies involving direct metallicities in \hii\ regions.

%In future studies, a robust study of how electron temperatures contribute to the ADF can be done to discover rigorous analytic relationships (i.e., Symbolic Regression) between strong emission lines and the ADF. Such a study would enable a deeper understanding of the underlying mechanism that leads to inhomogeneities inside of \hii\ regions.  

\acknowledgments

KK gratefully acknowledges funding from the Deutsche Forschungsgemeinschaft (DFG, German Research Foundation) in the form of an Emmy Noether Research Group (grant number KR4598/2-1, PI Kreckel) and the European Research Council’s starting grant ERC StG-101077573 (“ISM-METALS").

RSK acknowledges financial support from the ERC via Synergy Grant ``ECOGAL'' (project ID 855130) and from the German Excellence Strategy via the Heidelberg Cluster ``STRUCTURES'' (EXC 2181 - 390900948). In addition RSK is grateful for funding from the German Ministry for Economy and Energy (BMWE) in project ``MAINN'' (funding ID 50OO2206), and from DFG and ANR for project ``STARCLUSTERS'' (funding ID KL 1358/22-1). 

TGW gratefully acknowledges support from the UK ALMA Regional Centre (ARC) Node, which is supported by the Science and Technology Facilities Council grant number ST/Y004108/1.

Based on observations obtained at the Canada-France-Hawai`i Telescope (CFHT) which is operated by the National Research Council of Canada, the Institut National des Sciences de l'Univers of the Centre National de la Recherche Scientifique of France, and the University of Hawai`i. CFHT is located on Maunakea on Hawai`i Island, a mountain of considerable cultural, natural, and ecological significance. Maunakea is a sacred site to Native Hawaiians, also known as K\=anaka `\=Oiwi. We would like to thank the Canada-France-Hawai`i Telescope (CFHT) Operations and Software Groups for their contributions and diligence in maintaining observatory operations; the CFHT Astronomy Group for their observation coordination and data acquisition efforts; and the CFHT Finance \& Administration Group for their contributions to the management and administration of the observatory. Based on observations obtained with SITELLE, a joint project between Universit\'e Laval, ABB-Bomem, Universit\'e de Montr\'eal, and the CFHT with funding support from the Canada Foundation for Innovation (CFI), the National Sciences and Engineering Research Council of Canada (NSERC), Fonds de recherche du Qu\'ebec -- Nature et technologies (FRQNT) and CFHT.

Based on observations collected at the European Southern Observatory under ESO programmes 094.C-0623 (PI: Kreckel), 095.C-0473,  098.C-0484 (PI: Blanc), 1100.B-0651 (PHANGS-MUSE; PI: Schinnerer), as well as 094.B-0321 (MAGNUM; PI: Marconi), 099.B-0242, 0100.B-0116, 098.B-0551 (MAD; PI: Carollo) and 097.B-0640 (TIMER; PI: Gadotti).

Table \ref{tbl:sample_observations} includes distances that were compiled by \cite{Anand2021} from \cite{Freedman2001ApJ...553...47F} and \cite{Jacobs2009AJ....138..332J}. 
This research made use of Montage. It is funded by the National Science Foundation under Grant Number ACI-1440620, and was previously funded by the National Aeronautics and Space Administration's Earth Science Technology Office, Computation Technologies Project, under Cooperative Agreement Number NCC5-626 between NASA and the California Institute of Technology.

\bibliography{refs.bib}{} 

\clearpage

\appendix

\section{Background Continuum Extraction for Emission Line Fitting}\label{appendix:background} 

Each \hii\ region in our sample has a unique galactic and rotational velocity. Thus, the spectral ranges in which background continuum is extracted for emission line fitting varies for each \hii\ region in our sample. In the fitting algorithm designed in this study, a list of known atomic transitions \citep{NIST_elines} were used to designate where background continuum should be extracted in the rest frame wavelength. After the fitting process described in Section \ref{subsec:auroral} was carried out, a redshfited wavelength is measured and can be used to extract continuum background from spectral ranges that do not contain known atomic transitions. In Table \ref{tbl:back_continuum}, the wavelength ranges in which continuum background is extracted is shown. 

\begin{table*}[h!]
\centering
\caption{Wavelength Ranges for continuum background extraction for emission lines fitted in this work. The redshifted wavelength for a given emission feature is denoted with $\lambda$.}
\footnotesize
\centering
%\hspace*{-0.08\textwidth}
\begin{tabular}{ccccc}
\hline
\vspace{0.5ex}
\textbf{Emission Line} & $[\rm N\, II]\lambda$5755 & $[\rm S\, III]\lambda$6312 & $[\rm O\, II]\lambda$7320 & $[\rm O\, II]\lambda$7330 \\\hline
\vspace{0.5ex}
\textbf{Wavelength} & \([\lambda - 175 : \lambda - 8]\) & \([\lambda - 200 : \lambda - 70]\) & \([\lambda - 150 : \lambda - 10]\) & \([\lambda - 150 : \lambda - 20]\) \\
\textbf{Ranges [\AA]} & \([\lambda + 8 : \lambda + 100]\) & \([\lambda + 17 : \lambda + 45]\) & \([\lambda + 20 : \lambda + 150]\) & \([\lambda + 10 : \lambda + 150]\) \\
& \([\lambda + 180 : \lambda + 400]\) & \([\lambda + 63 : \lambda + 120]\) & & \\
\hline
\end{tabular}
\label{tbl:back_continuum}
\end{table*}

It should be noted that in NGC~4535 the spectral range $[\lambda+8]  \text{ --- }  [\lambda+100]$\AA\ is excluded for $[\rm N\, II]\lambda$5755. This region of the spectrum was removed in the MUSE DAP. NGC~4535 is the lone galaxy in this sample of 5 galaxies that made use of the ground-layer adaptive optics mode from MUSE. For this reason, spectral regions in NGC~4535 were masked due to a large quantity of sky lines. This part of the background continuum extraction is excluded in NGC~4535 to avoid underestimating the noise for the $[\rm N\, II]\lambda$5755 auroral line.

\end{document}